\documentclass[11pt]{article}
\usepackage{amsmath,amssymb}
\usepackage{graphicx,psfrag,epsf}
\pdfoutput=1 
\usepackage{enumitem}
\usepackage{natbib}
\usepackage{url} 
\usepackage{float}
\usepackage{caption}
\usepackage{rotating}
\usepackage{multirow}
\usepackage{accents}
\usepackage{pdflscape}
\usepackage[backref=page,colorlinks,citecolor=blue]{hyperref}
\usepackage[margin = 20pt]{subfig}
\usepackage{dcolumn}
\newcolumntype{d}[1]{D{.}{.}{#1}}
\usepackage[doublespacing]{setspace}

\usepackage{mathtools}

\usepackage{xcolor}
	\definecolor{BEALogoBlue}{RGB}{0, 76, 151}
	\definecolor{BEAlogoorange}{RGB}{216, 96, 24}
	\definecolor{BEAlogogray}{RGB}{158, 162, 162}
	\definecolor{BEAlightBlue}{RGB}{195, 215, 238}
	\definecolor{BEAmediumorange}{RGB}{242, 169, 0}
	\definecolor{BEAlightgray}{RGB}{220, 222, 223}
	\definecolor{barniepurple}{RGB}{128, 0, 128}

\usepackage{array}
\newcolumntype{L}[1]{>{\raggedright\arraybackslash}m{#1}}

\usepackage{tikz}
\usetikzlibrary{arrows.meta,positioning}
\usetikzlibrary{arrows}
\usetikzlibrary{shapes}
\usetikzlibrary{calc}
\usetikzlibrary{patterns}

\tikzstyle{decision} = [diamond, draw, fill=BEAlightBlue,
text width=3em, text badly centered, node distance=2cm, inner sep=0pt]
\tikzstyle{block} = [rectangle, draw, fill=BEAlightBlue, 
text width=6.25em, text centered, rounded corners, minimum height=2em]
\tikzstyle{block2} = [rectangle, draw, fill=BEAlightBlue, 
text width=10em, text centered, rounded corners, minimum height=2em]
\tikzstyle{line} = [draw, -latex']
\tikzstyle{cloud} = [draw, ellipse,fill=BEAlightgray, node distance=3cm,
minimum height=2em]
\tikzset{invisible/.style={rectangle, node distance=3cm},
	no arrow/.style={-, style={-}}}

\usepackage{booktabs}

\usepackage[noend]{algpseudocode}
\usepackage{algorithm}
\makeatletter
\def\BState{\State\hskip-\ALG@thistlm}
\makeatother

\usepackage{etoolbox}
\newcounter{bibcount}
\makeatletter
\patchcmd{\@lbibitem}{\item[}{\item[\hfil\stepcounter{bibcount}{\thebibcount.}}{}{}
\renewcommand\NAT@bibsetup%
[1]{\setlength{\leftmargin}{\bibhang}\setlength{\itemindent}{-\parindent}%
	\setlength{\itemsep}{\bibsep}\setlength{\parsep}{\z@}}
\makeatother

\newcommand{\blind}{0}
\begin{document}
	\def\spacingset#1{\renewcommand{\baselinestretch}%
		{#1}\small\normalsize} \spacingset{1}
	\if0\blind
	{
		\title{Standing on the Shoulders of Machine Learning: \\ Can We Improve Hypothesis Testing?\bf}
		\author{Gary Cornwall \thanks{The authors would like to acknowledge Marina Gindelsky, Scott Wentland, Tara Sinclair, Jeff Mills, and Olivier Parent for their helpful comments. The views expressed here are those of the authors and do not represent those of the U.S. Bureau of Economic Analysis or the U.S. Department of Commerce.
			}\hspace{.2cm}\\
			Bureau of Economic Analysis \\
			and \\
			Jeff Chen \\
			Bennett Institute for Public Policy\\
			University of Cambridge \\	
			and \\
			Beau Sauley \\
			University of  Cincinnati}
	
		\maketitle
	} \fi
	\if1\blind
	{
		\bigskip
		\bigskip
		\bigskip
		\begin{center}
			{\LARGE\bf Standing on the Shoulders of Machine Learning: \\ Can We Improve Hypothesis Testing?}
		\end{center}
		\medskip
	} \fi
	\bigskip
	\begin{abstract}
			 In this paper we have updated the hypothesis testing framework by drawing upon modern computational power and classification models from machine learning. We show that a simple classification algorithm such as a boosted decision stump can be used to fully recover the full size-power trade-off for any single test statistic. This recovery implies an equivalence, under certain conditions, between the basic building block of modern machine learning and hypothesis testing. Second, we show that more complex algorithms such as the random forest and gradient boosted machine can serve as mapping functions in place of the traditional null distribution. This allows for multiple test statistics and other information to be evaluated simultaneously and thus form a pseudo-composite hypothesis test. Moreover, we show how practitioners can make explicit the relative costs of Type I and Type II errors to contextualize the test into a specific decision framework. To illustrate this approach we revisit the case of testing for unit roots, a difficult problem in time series econometrics for which existing tests are known to exhibit low power. Using a simulation framework common to the literature we show that this approach can improve upon overall accuracy of the traditional unit root test(s) by seventeen percentage points, and the sensitivity by thirty six percentage points. 
	\end{abstract}
	\noindent%
	\vfill
	\newpage
	

\section{Introduction}
	This paper re-examines hypothesis testing through the lens of 21$^{st}$ century computational capabilities and machine learning methods. For at least a century, hypothesis testing has been a basic, fundamental pillar of statistical inference and the analysis of data. Yet, while statistical methods used in academic research are perpetually evolving, many of the hypothesis tests commonly used today have not been substantially improved in decades. We revisit one such test, the unit root test, as a specific application that we exploit in an effort to illustrate a more general approach to deriving hypothesis test statistics using machine learning methods. In this application, we find that our approach can improve upon the traditional unit root test(s) by seventeen percentage points over the best single-test alternative. The majority of this improvement comes from the ability to more accurately reject the null of a unit root when the null is false, as our method shows a thirty-six percentage point increase in sensitivity (power) relative to the best single-test alternative. This improvement, while significant for unit root tests in its own right, also provides initial evidence that traditional approaches to hypothesis testing can be re-imagined using machine learning methods in ways that were not computationally tractable a century ago (but perhaps better aligns with the original intent of \cite{neyman1933ix} and the early development of hypothesis testing).
	
	The motivation of this study comes not from the context of unit roots, but rather from origins of the modern-day hypothesis test structure. Every problem in statistical inference, no matter how mundane, comes with some degree of computational difficulty. In the early twentieth century, when the current hypothesis testing framework was developed, the cost of computation was astronomically high by today's standards \citep{nordhaus2007two}. Ultimately, this shaped the process of deriving test statistics. The prevailing approach was to fully characterize a null distribution and derive decision rules under conditions where the computational costs were paid up front by a group of ``human calculators" rather than by the researchers on demand. The result is that nearly every statistics text book published contains appendices which provide statistical tables for various distributions used by test statistics. Today, it is still common to see, perhaps by convention rather than computational necessity, the development of new statistics that continue to leverage this paradigm, proving either that the test statistic is distributed under some known form (\textit{e.g.,} a standard normal distribution) or providing an analytic solution to the distribution of the assumed null.
	
	The null distribution serves as the backdrop to the mapping function for the hypothesis space, and is used to identify an appropriate threshold (\textit{e.g.,} $0.05$) upon which the decision rule is based. Yet, implicit in the traditional approach is the notion that the distribution of the alternative is not a central concern, primarily because it often has no tractable solution. This has led to a paradigm where our decision is based upon static Type I errors, typically denoted by $\alpha$, which is defined as incorrectly rejecting the null when it is true. One historical artifact from the discussions between R.A. Fisher and Neyman-Pearson in the early twentieth century is our conventional choice of $\alpha \in \{0.10, 0.05, 0.01\}$ which stemmed from Fisher's insistence upon $5\%$ as the standard \citep{lehmann1993fisher}.\footnote{The battles between R.A. Fisher and Neyman-Pearson over this topic are often discussed as being both acrimonious between the parties and downright damaging to the profession (see \cite{zabell1992ra} as an example).} However, this standard does not guarantee optimal accuracy, some objective trade-off, rather it only guarantees that the uniformly most powerful test will be the one which minimizes Type II errors conditional upon a bounded Type I error rate \citep{neyman1933ix}. Indeed, it was never the intention of \cite{neyman1933testing} that a fixed Type I error rate be chosen. Rather they wrote: ``It is also evident that in certain cases we attempt to adjust the balance between the risks $P_1$ and $P_{II}$ to meet the type of problem before us." (\citealp*{neyman1933testing}, pp. 497). Thus, one objective of this paper is to re-imagine hypothesis testing as \cite{neyman1933testing} might have if they had access to the computational tools we have at our disposal today.
	
	Rather than choosing an arbitrary, fixed Type I error, the machine learning approach put forth herein chooses a threshold that directly reflects the cost or loss function, a process which is more in line with the original intent of hypothesis testing. One key advantage of examining a hypothesis through a model-based approach such as boosting, decision trees, and other machine learning algorithms is the ability to identify, based on the training set, thresholds of classification which represent a different weighting structure to the error types. More generally, we recognize that this approach diverges from the well-established asymptotic traditions of modern hypothesis testing by taking arbitrary thresholds as given. However, we view this departure as a feature, and a key contribution to the hypothesis testing literature that leads to better test statistics and a deeper understanding of relevant trade-offs. In Section \ref{sec: mltest} we will outline this with respect to tests for unit roots and show for example that the typical $5\%$ critical value for the Augmented Dickey-Fuller test represents a case where the cost of a Type II error, $c(e_2)$ is two to three times the cost of a Type I error, $c(e_1)$. While we will expand upon the importance of a unit root test in Section \ref{sec: unitroots}, it seems likely that this cost structure is less than ideal given the consequences of failing to identify a unit root.
	
	Given a set of null and alternative hypotheses, we propose that machine learning classifiers can improve the quality of hypothesis tests by serving as the mapping functions. We find that under certain conditions, the test statistic mechanism is equivalent to a ``weak learner", which is defined as a learner which, when classifying an object, will produce results that are slightly better than random chance. Furthermore, more powerful machine learning techniques such as boosting \citep{schapire2013boosting} can recover the full size-power trade-off for the test in question. For many cases, especially for those tests which are deemed pivotal to subsequent modeling decisions and/or inferences, there often exists a large portfolio of test statistics available to researchers (\textit{e.g.} tests for a unit root in time series econometrics). Having shown that a hypothesis test is a weak learner, and drawing inspiration from the ensemble forecasting literature (see \cite{elliott2004optimal,timmermann2006forecast} as examples), we use more complex machine learning classification models such as the random forest \citep{breiman2001random} and gradient boosting machines \cite{chen2016xgboost} to exploit unused variation between the test statistics. Moreover, metrics commonly used to judge classifiers in the machine learning literature such as the Matthews Correlation Coefficient and F-measure \citep{hastie2009elements,kuhn2013applied,tharwat2018classification} show the approached contained herein as strictly dominating traditional hypothesis tests for unit roots. This difference in efficacy is supported by evaluating the Receiver Operating Characteristic curves and more traditional power analysis.
	
	In this study, we limit the scope of our application to unit roots, which is an important data element in time-series econometrics. Unit root tests are commonly used in time-series analysis because the presence of a unit root may create spurious or highly misleading inferences, depending on the context \cite{granger1974spurious}. On the other hand, eliminating all unit roots can destroy relationships which are important to researchers for series which are co-integrated \citep{granger1981some, engle1987co}. This dichotomy is what makes test for unit roots so important for applied time-series research and has catalyzed over four decades of research into test statistics useful under a variety of assumed data generating processes.
	
	An additional benefit of this particular application is that, as statistical tests are defined primarily by their accuracy, we are able to evaluate the current state of testing performance using simulation designs that are already common in the unit root literature. Not only will this more model-based approach to hypothesis testing afford us the ability to explicitly incorporate cost into the choice of threshold, we will also gain the ability to exploit variation across a suite of test statistics. In many cases subsequent research has taken an original concept and developed new test statistics which relax assumptions, change the object of measurement, or seek to make an existing statistic more robust. Oftentimes the new test has some attractive feature that improves upon the original design, \textit{e.g.} more power at a lower number of observations. However, in some cases, this tinkering means the two tests can disagree on the disposition of the null as we will show in Section \ref{sec: mltest}. This happens because classification from the first test is unlikely to be a perfect subset of the second. This means that, depending on which test a practitioner uses, there may be a reasonable disagreement to be had regarding that aspect of the data. This reasonable disagreement can be disastrous for  not only the practitioner, but also the scientific community which builds upon that result. We will show in Section \ref{sec: mltest} that there is a great deal of variation between these test statistics which means that this ``reasonable disagreement" is likely to occur between tests and potentially practitioners or researchers. As a result, the conventional wisdom has become ``When in doubt, assume a unit root." which may or may not be appropriate depending on the context.
	
	The remainder of this paper is structured as follows. In Section \ref{sec: unitroots} we describe the unit root problem and analyze the accuracy of current hypothesis tests. In Section \ref{sec: bridging} we bridge hypothesis testing with binary classification problems that are common in machine learning and show that modern boosting algorithms can recover the full Receiver Operating Characteristic curve of the test statistic. The use of these algorithms point to more accurate results which are not constrained to a fixed Type I error rate. Section \ref{sec: mltest} lays out an ML-based unit root test that draws on ensemble machine learning methods and delivers marked improvement in test accuracy. Section \ref{sec: beaugoestotown} applies these advancements to macroeconomic time series that have long been debated in the literature. Section \ref{sec: conclusion} concludes and offers avenues for additional research.
	
\section{Unit Roots and The Current State of Hypothesis Testing}\label{sec: unitroots}

	\subsection{Why Unit Roots?}  \label{unitrootsdgp}
		Of the statistical tests in time-series econometrics perhaps none are more visible, and has attracted as much research over the previous five decades, as that of the unit root. Its importance is made plainly obvious by \cite{granger1974spurious} which outlines the concept of spurious regression, showing that regressions involving random walks can create relationships between otherwise independent series. Subsequently, a great deal of research capital has been spent on developing and improving tests for unit roots and stationarity with some of the more well-known including \cite{dickey1981likelihood}, \cite{phillips1988testing}, \cite{kwiatkowski1992testing}, \cite{elliot1996efficient}, \cite{ng2001lag}, and \cite{bai2004panic} among many others.
		
		Herein we will focus on three unit root structures that are used throughout the related literature. Let $y_t$ be a autoregressive time series generated such that one of the following is true,
		\begin{align}
			y_t &= \lambda + \phi y_{t-1} + \delta t + \epsilon_t \label{eq: dpg 1}\\
			y_t &= \lambda + \phi y_{t-1} + \epsilon_t \label{eq: dpg 2}\\
			y_t &= \phi y_{t-1} + \epsilon_t, \label{eq: dpg 3}
		\end{align}
		indexed by $t = 1,\ldots,T$. In Equation \ref{eq: dpg 1} we have included a drift term, $\lambda$, as well as a time trend, $\delta t$, and in all cases $\epsilon_t \sim N(0,\sigma^2)$. Equations \ref{eq: dpg 2} and \ref{eq: dpg 3} omit the time trend and both the time trend and drift term respectively. 	 
		
		Equation \ref{eq: dpg 3} is the simplest case and often is the first used to introduce the concept of unit roots in econometric texts. Alternatively one could write this process using the lag operator as $(1-\phi L)y_t = \epsilon_t$ such that $Ly_t = y_{t-1}$. We will assume that $\epsilon_t \sim \text{N}(0,\sigma_t^2) \ \forall \ t$ and that $\sigma^2_1 = \ldots = \sigma^2_T$. The polynomial, $(1 - \phi L)$ has a root of $1/\phi$ and if $|\phi|<1$ then $y_t$ is considered stationary. This process is one which does not depend on $t$ and over the long run has a vanishing memory. If $\phi = 1$ then the series $y_t$ is a random walk and contains a unit root, a potentially dangerous attribute that needs to be identified. 
		
		When transformed into a difference equation this leads to a hypothesis whereby the null is $H_0: \phi = 1$ and the alternative is $H_1: |\phi| < 1$. This structure is used in \cite{dickey1981likelihood}, \cite{phillips1988testing}, and \cite{elliot1996efficient} among others. Since many of these tests are effective in  rejecting the null when $|\phi| \rightarrow 0$ we will focus our efforts on a very narrow subset of the domain such that our null remains $H_0: \phi = 1$ while the alternative is $H_1: |\phi| \in \{.90000, .99999\}$. It is at this interface between unit roots (UR) from \textit{near} unit roots (NUR) that a test's worth is proven.
		
		Testing for a unit root requires one to make two decisions. First, one must choose a data generating process which best represents the data. This choice dictates the distribution of the null and resulting critical values for the test itself. It should be apparent that there is ample room for human error -- if alternative structure is a better approximation of the true data generating process, then a practitioner can make an error by choosing the wrong framework for the null distribution. Second, once a test scenario is established the practitioner must choose some level of long-run type I error rate, $\alpha$, they are willing to accept with the most common choices $\alpha \in \{0.01, 0.05, 0.10\}$. This two-choice framework leads to intra-test variation, the choice of Equation \ref{eq: dpg 1} with $\alpha_1 = 0.05\%$ may lead to a different result than using Equation $\ref{eq: dpg 2}$ with $\alpha_2 = 0.05\%$. 	Moreover, depending on the selected test (\textit{e.g.} Augmented Dickey-Fuller versus Phillips-Perron) there can be between-test variation -- that is one test may reject the null while the other fails-to-reject it.
		
		We exploit this variation by employing classification algorithms as mapping functions in place of the traditional null distribution. Additionally, by limiting design to the three cases outlined above we have a clean simulated environment that is well known to the unit root literature and allows us to evaluate not only our classification results but ensure the data being generated fulfills some, if not all, of the null assumptions. While there are a bevy of tests to choose from, we will use nine different tests for unit roots, many -- but not all -- of which have the structure outlined above with respect to the null and alternative. These tests include the ADF \citep{dickey1981likelihood}, PP \citep{phillips1988testing}, KPSS \citep{kwiatkowski1992testing}, PGFF \citep{pantula1994comparison}, Breit \citep{breitung2002nonparametric,breitung2003corrigendum}, ERS-d and ERS-p \cite{elliot1996efficient}, URSP \citep{schmidt1992lm}, and URZA \citep{zivot2002further}.\footnote{The majority of these tests are implemented in the ``urca" and ``egma" packages in R. We use these packages in combination with R version 4.0.2 ``Taking Off Again" for all subsequent simulations and calculations.} 
		
	\subsection{The Structure of Hypothesis Tests}
		At the most basic level, a hypothesis test is nothing more than rules which govern statements about a population parameter \citep{casella2002statistical}. Typically, these tests are comprised of three components:
		\begin{enumerate}[noitemsep]
			\item A pair of complementary hypotheses; 
			\item A mapping function that translates input parameters into the hypotheses space; and
			\item A decision threshold.
		\end{enumerate}
		
		The first of these components -- the complementary hypotheses -- frames the extent of a problem space so that one can make a claim that one of the two hypotheses is true, typically drawing on a sample from the population of interest. Generally, we consider the following format in modern testing,
		\begin{align}
		H_0: \theta \in \Theta_0 \\
		H_1: \theta \in \Theta_0^c
		\end{align}
		where $\theta$ is a sample estimate of interest, $\Theta_0 \cup \Theta_0^c \in \Theta$, and $\Theta$ represents the parameter space. For simplicity, we will rely on the common nomenclature and label $H_0$ as the null hypothesis and $H_1$ as the alternative. In this context it is assumed that $H_0$ is true unless evidence from the sample is sufficient to reject, hence the often times maddening language of \textit{reject} and \textit{fail to reject}.  
		
		While the hypothesis is effectively the same from one unit root test to the next, the underlying empirical machinery tends to differ. For this, we turn our attention to the mapping function and decision threshold. The mapping function is comprised of a test statistic -- which is a function of the sample and thus a random variable itself -- as well as the null distribution used to assess the support for $H_0$. The score produced by the mapping function is translated by way of a decision threshold that forces a verdict of reject or fail to reject the null. A test's accuracy is dictated by the rejection threshold chosen to fix the long-run Type I error rate, $\alpha$.
		
		These three components can be seen in the [Augmented] Dickey-Fuller test -- one of the most relied upon tests for unit roots.  The original test statistic can be expressed as,		
		\begin{align}
		\hat\tau = (\hat\phi - 1)S^{-1}_e(\sum_{t = 2}^{N}Y^2_{t-1})^{1/2},
		\end{align}
		where,
		\begin{align}
		S^{-1}_e = (n - 2)^{-1}\sum_{t = 2}^{N}(Y_t - \hat\phi Y_{t-1})^2,
		\end{align}
		with limiting distribution derived in \cite{dickey1979distribution}.\footnote{While we outline the Dickey-Fuller test here, going forward we will use the Augmented Dickey-Fuller test which adds additional lags of the observed time series.} The null hypothesis of this test, $H_0: \rho = 1$, is well-known and has spawned years of follow-up research. Most commonly this test is performed on a model written as,
		\begin{align}
		\Delta y_t = \gamma y_{t-1} + \epsilon_t
		\end{align}
		where $\Delta y_t = y_{t} - y_{t-1}$ is the first difference. This means that testing the null $H_0: \gamma = 0$ is equivalent to testing $H_0: \phi = 1$ from Equation \ref{eq: dpg 3}. Since $|\phi| \in (0,1)$ and the first difference implies $\gamma = 1 - \phi$ the alternative is $H_1: \gamma < 0$ rather than the traditional $H_1: \gamma \neq 0$ used for regression parameters. The limiting distribution does not have a convenient closed form expression and as a result the critical values for common levels of $\alpha$ have been derived through Monte Carlo simulations (\textit{e.g.} $\alpha_{0.05} = -1.95$ for the simplest case \cite{Hamilton1994}).

\subsection{Accuracy of Current Tests} \label{test-accuracy}
	
	Simulation studies have become rather ubiquitous in the unit root literature, in part because the null distribution for many of the tests has no analytic form. As a result, critical values are often obtained through simulation (see \cite{mackinnon2010critical} for one such example) and are thus dependent on the assumptions applied to the simulation environments under which they were constructed. In order to assess the performance of current unit root tests, we simulate a large number of time series, apply tests to these simulations, then compare their ability to detect unit roots.
	
	To that end we will generate $M = 500,000$ series, of which $350,000$ and $75,000$ are reserved for training and validating ML algorithms, respectively. The remaining $75,000$ series will serve as the basis of accuracy comparisons for all subsequent sections.  To begin let $\pi_u \sim U(0,1)$ such that,
	
	\begin{align}
		U = \begin{cases}
			+ 1: \ \pi_u \geq 0.50 \\
			- 1: \ \pi_u < 0.50
		\end{cases},
	\end{align}

	where $U = +1$ denotes a series with an Unit Root and $U = -1$ a Near Unit Root. Conditional upon the series containing a Unit Root we generate data from Equation \ref{eq: dpg 1} , \ref{eq: dpg 2} , and \ref{eq: dpg 3} where $\phi = 1$. If $U = -1$ then we draw $\phi$ uniformly over the interval $(0.9000,0.9999)$.\footnote{We have examined other distributions which govern the draw of $\phi$ in the case of a Near Unit Root including $U(0.00,0.99)$, $B(\alpha = 8, \beta = 1)$, and $B(\alpha = 2,\beta = 2)$. We have also varied relative weights given to Unit Roots and Near Unit Roots by changing the cutoff for $\pi_u$. All additional results and code will be available on our website.} For all simulated series $\epsilon_t \sim N(0,1)$.\footnote{We have also varied the distribution of the error term by using $U(-1,1)$ with similar results.} Since the frequency of a series is orthogonal to its unit root status we assume all series are of monthly frequency.\footnote{We have varied the frequency in earlier iterations of this process but have left it as monthly in order to limit parameters in the simulated environment.} Each series is variable in length with the number of years drawn uniformly over the interval $(5,50)$. From each series we calculate the appropriate test statistic (\textit{e.g.} Augmented Dickey-Fuller) and record the test's conclusion assuming $\alpha = 0.05$ for all tests.\footnote{If given a choice of lag lengths for model selection, such as in the case of the Augmented Dickey-Fuller statistic, we choose based on Schwarz Information Criterion (also known as Bayesian Information Criterion  or BIC) \cite{schwarz1978estimating}.}
	
		\begin{table}[H]
			\centering
			\caption{A $2\times2$ Confusion Matrix}
			\label{tbl: confuse}
			\begin{tabular}{l|c|c}
				\toprule
				&        True NUR        &        True UR         \\ \midrule
				Predicted NUR & \small{True Positive (TP)}  & \small{False Positive (FP)} \\ \midrule
				Predicted UR  & \small{False Negative (FN)} & \small{True Negative (TN)}  \\ \bottomrule
			\end{tabular}
		\end{table}

	To evaluate efficacy of each test, we compare its performance based on standard measures from the machine learning literature, all of which are calculable from a $2 \times 2$ confusion matrix  \citep{hastie2009elements,kuhn2013applied, tharwat2018classification}. As outlined in Table \ref{tbl: confuse}, a confusion matrix produces the basic building blocks for a rich assortment of accuracy measures that emphasize different concepts, such as accuracy  (ACC), sensitivity (SEN), specificity (SPE), positive predictive value (PPV), negative predictive value (NPV), F-Measure (various weights), and the Matthew's Correlation Coefficient (MCC). Each measure allows a practitioner to optimize a model's performance for qualities that are pertinent to the application at hand. In Table \ref{tbl: efficacy calc} we layout the  calculations for each of these measures. In this case, as we will assume going forward, a series with a unit root in truth is a ``positive case", thus a true positive implies that the test identifies a series with a unit root as a series with a unit root.
	

		\begin{table}[H]
			\centering
			\caption{Calculating Measures of Efficacy}
			\label{tbl: efficacy calc}
			\begin{tabular}{l|c}
				\toprule
				Measure                                    &                                    Calculation                                    \\ \midrule
				\vspace{.5em}
				Accuracy                  &                      $\frac{(TP + TN)}{(TP + TN + FP + FN)}$                      \\
				\vspace{.5em}
				Sensitivity               &                              $\frac{TP}{(TP + FN)}$                               \\
				\vspace{.5em}
				Specificity               &                              $\frac{TN}{(TN + FP)}$                               \\
				\vspace{.5em}
				Positive Predictive Value &                              $\frac{TP}{(TP + FP)}$                               \\
				\vspace{.5em}
				Negative Predictive Value &                              $\frac{TN}{(TN + FN)}$                               \\
				\vspace{.5em}
				F-Measure                 &        $\frac{(1 + \beta^2)(TP)}{((1 + \beta^2)(TP) + \beta^2(FN) + FP)}$         \\
				\vspace{.5em}
				MCC                       & $\frac{(TP \times TN) - (FP \times FN)}{\sqrt{(TP + FP)(TP + FN)(TN+FP)(TN+FN)}}$ \\ \bottomrule
			\end{tabular}
		\end{table}

	Among the most commonly evaluated are sensitivity (also known as recall) which is a measure of positive accuracy relative to the Type II errors, and specificity, which is a measure of negative accuracy relative to the Type I errors.\footnote{In fact, it is trivial to show that specificity, in the context of a hypothesis test is equal to $1-\alpha$ where $\alpha \in \{0.01,0.05,0.10\}$ is the long-run Type I error rate typically chosen by researchers in the Neyman-Pearson framework.} In short, a high sensitivity implies a low number of Type II errors, and a high specificity implies a low number of Type I errors. Both measures are over the interval $(0,1)$ with a ``perfect" classifier resulting in both a sensitivity and specificity of $1$. Positive predicted value, also known as \textit{precision}, is a measure of how many predicted positives were in fact true positives. The F-Measure, $F_{\beta}$, is the weighted harmonic mean of both PPV and sensitivity, with weights determined by $\beta$. A value of $\beta < 1$ indicates that more emphasis is placed on precision, while a value of $\beta > 1$ favors sensitivity. For our purposes we will assume $\beta = 1$ and thus we place equal weight upon both precision and sensitivity. An $F_\beta$ of $1$ means that a classifier has perfect sensitivity and precision. Finally, the MCC \citep{matthews1975comparison, tharwat2018classification} is a measure of correlation between the true and predicted classifications and lies in the interval $(-1,1)$. A value of $1$ indicates perfect prediction while a value of $-1$ corresponds to total disagreement; an MCC of zero is random selection. Often times the MCC is considered to be a superior measure of classification ability since it is relatively invariant to changes in the data distribution.

	When the battery of nine unit root tests are applied to the $M = 75,000$  series, we observe clear strengths and weaknesses --  suggesting that before one places trust on any test, its qualities should be carefully examined. As is apparent in Table \ref{tbl: eval_res}, the ADF, ERS-d and ERS-p tests achieve the highest performance, achieving accuracy over 76\%.  While convenient for narrative purposes, overall accuracy obscures the realities of a test's performance, thus one needs to evaluate each of the classification measures.  The sensitivity (\textit{i.e.} true positive rate) for these tests is less than 60\% while the specificity  (\textit{i.e.} true negative rate) is over 90\%, indicating that current tests are more likely to identify a unit root correctly than a near unit root. This lower performance for near unit roots points to ``leakage" in the tests -- current tests incorrectly classify near unit roots as unit roots to a large degree. This is consistent with previous research showing tests for unit roots tend to have low power \citep{kennedy2008guide} and as a result higher sensitivity would be preferable.
	
	\begin{table}[H]

\caption{Accuracy Evaluation of Current Unit Root Tests }
\label{tbl: eval_res}
\centering
			\scalebox{.85}{
\resizebox{\linewidth}{!}{
\begin{tabular}[t]{lccccccc}
\toprule
  & ACC & SEN & SPE & PPV & NPV & F$^1$ & MCC\\
\midrule
ADF & 0.763 & 0.546 & 0.980 & 0.964 & 0.684 & 0.697 & 0.583\\
PP & 0.744 & 0.512 & 0.975 & 0.953 & 0.667 & 0.666 & 0.549\\
KPSS & 0.614 & 0.250 & 0.977 & 0.916 & 0.567 & 0.393 & 0.331\\
PGFF & 0.745 & 0.499 & 0.989 & 0.978 & 0.665 & 0.661 & 0.560\\
BREIT & 0.672 & 0.361 & 0.981 & 0.951 & 0.607 & 0.524 & 0.437\\
ERSd & 0.762 & 0.545 & 0.979 & 0.963 & 0.683 & 0.696 & 0.582\\
ERSp & 0.770 & 0.564 & 0.976 & 0.958 & 0.692 & 0.710 & 0.592\\
URZA & 0.635 & 0.309 & 0.959 & 0.883 & 0.582 & 0.458 & 0.354\\
URSP & 0.727 & 0.552 & 0.903 & 0.850 & 0.669 & 0.669 & 0.485\\
\bottomrule
\end{tabular}}
}
\caption*{\footnotesize{\textit{Note: }In this table we provide common measures of efficacy derived from the $2 \times 2$ confusion matrix for existing tests. In this case we are evaluating the $75,000$ series we later use as a test set. This includes data from all three of the aforementioned DGPs and all tests were assessed using $\alpha = 0.05$. }}
\end{table}

	The quality of tests vary across different conditions, such as values of $\phi$ and the sample size. We can measure tests' power -- in this case, equivalent to sensitivity  --  by comparing the empirical power. The ADF test, conditional upon the DGP outlined in Equation $\ref{eq: dpg 3}$, is among the most powerful test statistics over the simulated sample ($45\%$ sensitivity -- see Table \ref{tbl: enders3} in the Appendix). In Figure \ref{fig: power} we have plotted traditional power curves for four of the unit root tests under examination. Here we have slightly expanded the interval of $\phi$ such that $\phi \in (.85,1)$, moved in steps of $.005$, and completed $20,000$ iterations at each step. Data was generated from Equation \ref{eq: dpg 3} directly with $T \in \{25,50,100,250\}$, $\epsilon_t \sim N(0,1)$, and $y_1 = 0$.  As may be expected, test performance tends to be weak in small samples (\textit{e.g.} $T=25$ does not exceed Power of $0.25$) and grows with more degrees of freedom. While these figures are supportive of the ADF test as the more powerful option of the four, though it is helpful to remember that each test is often designed with some slight differences in mind. For example, some tests are focused on achieving high performance in the presence of a moving-average term rather than the DGP we have specifically used.

	\begin{figure}[H]
		\centering
		\caption{Tests for a Unit Root Have Low Power} \label{fig: power}
		\subfloat{
			\label{fig: power25}
			\includegraphics[width = 1\linewidth, keepaspectratio]{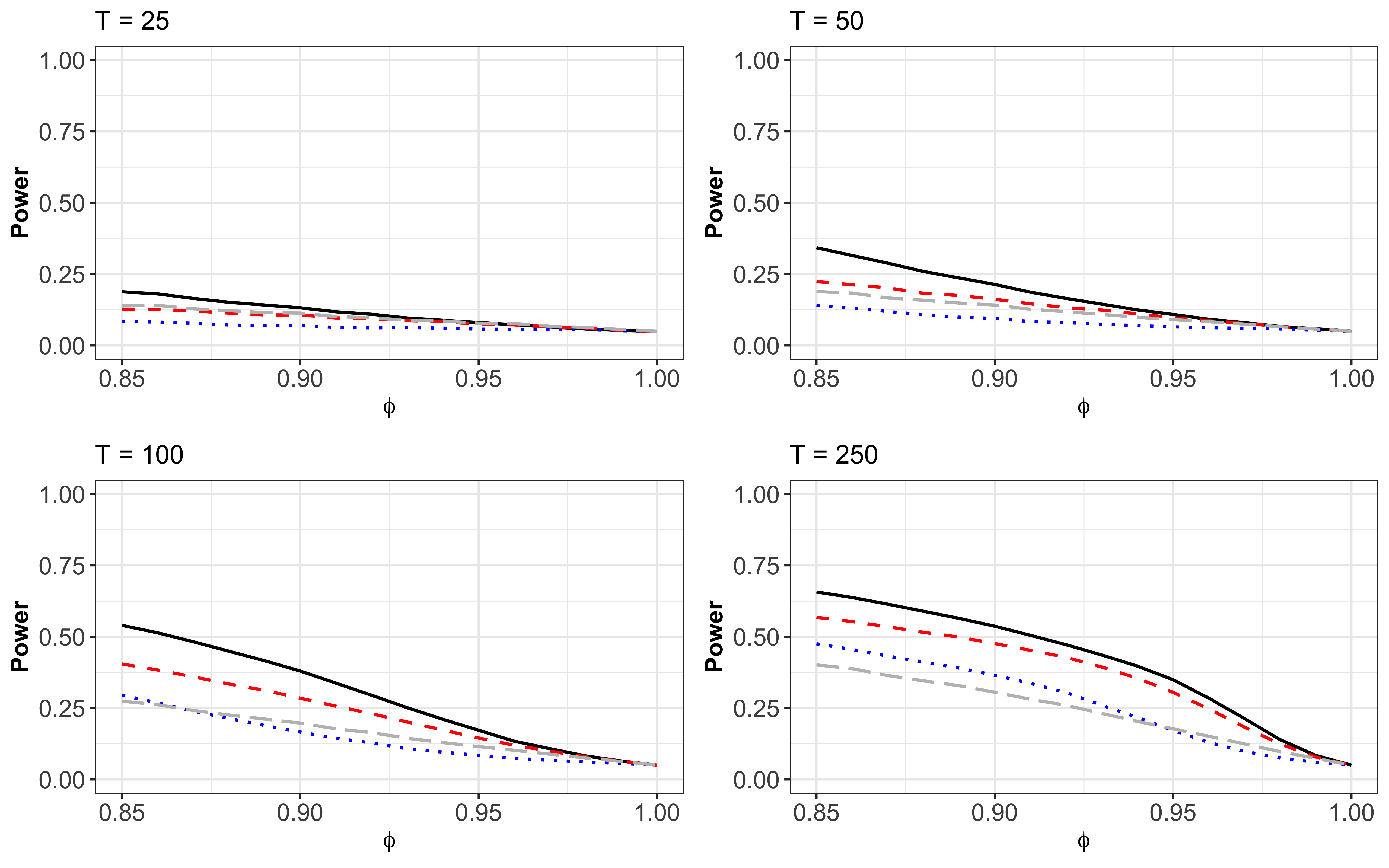}
		} 
		\caption*{\footnotesize{\textbf{Note: }This figure provides the traditional power curves conditional upon $T \in \{25,50,100,250\}$ for the ADF (black solid line), PP (blue dotted), ERS (red dashed), and Breit (grey long-dash) tests. For simplicity, we focus only on the DGP in Equation \ref{eq: dpg 3} and have identified empirical critical values that allow the tests to be comparable at $\alpha = 0.05$. Note that in Table \ref{tbl: efficacy calc} we have an assortment of series lengths, $T \in (60,600)$. Moreover, it is important to remember that some of these statistics, as mentioned previously, were designed to optimize power under slightly different conditions (\textit{e.g.} heteroskedastic error terms). }}
	\end{figure}
	
	One other piece of insight which can be drawn from Figure \ref{fig: power} and Table \ref{tbl: efficacy calc} is that there must be some variation in labeling between test statistics. That is, despite the lower power of the Phillips-Perron test under this DGP, in smaller samples, it is unlikely that the set of series identified as containing a unit root by the Phillips-Perron test is a perfect subset of the Augmented Dickey-Fuller test. This same argument could be made for the other tests as well. If this is the case, and we will show you that it is shortly, then using any single test will leave information on the table. Exploiting this between-test variation is one of the main cases for using classification algorithms such as the random forest and gradient boosting as mapping functions in place of the traditional null hypothesis, a subject we will delve into more thoroughly in Section \ref{sec: mltest}.
	
\section{Bridging Hypothesis Tests with Binary Classification}\label{sec: bridging}
	
	In this section we will provide details on the most basic of weak learner (the decision stump) and how it can choose a threshold for decision over a continuous support. We will then show that for some $\alpha = \alpha'$ the chosen thresholds, ($x_\alpha \equiv x_0$), are equivalent and thus a hypothesis test, $g(x)$, is equivalent to a decision stump, $h(x)$, conditional upon $\alpha = \alpha'$. We will  also show that when weak learners are combined into a strong learner, such as in the case of Adaptive Boosting \citep{schapire2013boosting}, it is possible to fully recover the power-size trade off of a hypothesis test. In other words, a boosted decision stump is equivalent to a hypothesis test.

	\subsection{Connecting the Dots}
	
	Hypothesis tests have all the trappings of a binary classification problem. The typical classification model is developed and applied in three steps:  \textit{training, validation, and testing}. In the training step, a model learns from a sample of input features (i.e. right-hand side variables) in order to classify instances by one of two classes of a binary target (i.e. dependent variable). For example, let $\mathcal{D} = \{(x_1,y_1),\ldots,(x_M,y_M)\}$ be a set of training data, indexed by $m = (1,\ldots,M)$. It is assumed these observations are independent and identically distributed, with each $x_m$ a realization of random variable $\mathcal{X}$ having support $\mathbb{R}$.\footnote{Note that for simplicity we are assuming that there is but one observed $x$ for each observation. It is trivial to assume that $\boldmath{x}_m$ is a $K$-dimensional vector of observed features such that $X \in \mathcal{X} \subset \mathbb{R}^K$.} Following convention we will denote the collection of observations as a vector with the capital such that $\mathcal{D} = (X,Y)$. A classifier, $h$, is a mapping, $h: \mathcal{X} \rightarrow \{-1,+1\}$, which returns the predicted class, $y$, for each $m$, conditional upon $x$. For example, we can consider the following classifier,	
	\begin{align}
		y_m = \begin{cases}
			+1 \ \mathrm {if} \ x_m \in \zeta \\
			-1 \ \mathrm {if} \ x_m \in \zeta^c
		\end{cases},
	\end{align} 
	where $\zeta \subset \mathcal{X}$ such that $\zeta \cup \zeta^c = \mathcal{X}$.\footnote{The use of a binary classification system is merely for its simplicity and readability. In practice $y \in Y$ where $Y$ has some finite cardinality, $k$.} Both simple and complex classifiers can partition $\mathcal{D}$ into classifications $y = +1$ and $y = -1$ conditional upon $\mathcal{X}$.  Interestingly, this is nearly identical to the definition of a "statistical test" as  provided by \cite{neyman1933testing}.
	
	The  objective of the validation step is to evaluate the performance of candidate models on an unbiased data set that is independent of the training sample. Model validation designs can take on any number of forms. A simple design involves splitting a sample into three randomly assigned partitions -- a training set for fitting a model, a validation set to identify the best candidate model, and a test set to evaluate the best model's performance. In each step, the model validation design ensures identical and independently distributed samples, thereby eliminating bias in performance estimates. The design can also be evolved to meet the specific requirements of the application at hand. For example, the training sample could serve as its own validation sample through k-folds cross validation, allowing the validation set to be used to calibrate a decision threshold that reflects preferred relative costs between Type I and II errors (e.g. NP testing requires an explicit choice of error weights), and the test sample remains as the final evaluation. 
	
	Only once a model achieves a desired level of accuracy when applied to the out-of-sample test can it advance to the second stage of the classification process-- \textit{scoring}. A model is said to score new, previously unseen instances by mapping them to the hypothesis space, thereby resulting in a probability of belonging to the positive class ($Pr(y_m = +1 | X)$). To convert a probability into a prediction of class membership requires one to make an explicit choice of a classification threshold. In a balanced two-class classifier, a threshold can be selected to explicitly reflect the Type I and Type II costs. With this in mind, we can consider traditional hypothesis tests, such as the ADF and KPSS, as effectively simple binary classification scoring models. Given the striking resemblance, we map the similarities between hypothesis tests and classification problems in Table \ref{tbl: comparehyp}.
	

		\begin{table}[H]
			\centering
			\caption{Drawing parallels between hypothesis tests and classification models}
			\begin{tabular}{lcc}
				\toprule
				Qualities &  Hypothesis Tests   &        Classification problem      \\ \midrule
				Outcomes & Complementary hypotheses & Binary target    \\ 
				Mapping function &  Test statistic and null distribution &  Classification algorithm  \\ 
				Model Validation &  Type I and Type II errors &  Type I and Type II errors  \\ 
				Threshold & Rejection threshold&  Decision threshold  \\ \bottomrule
			\end{tabular}
		\end{table}\label{tbl: comparehyp}

	It is clear that both classification problems, as outlined in the machine learning literature, and hypothesis tests share a great deal in common. The primary differences lie in the mapping function and the resulting choice of threshold. Let us begin with perhaps the simplest learning algorithm, the decision stump. A decision stump can be thought of as a severely pruned decision tree \citep{oliver1994averaging}, with its purpose to minimize the risk function $R(h) = \mathbb{P}(Y = -1)R_1(h) + \mathbb{P}(Y = +1)R_2(h)$ where $R_1(h) = \mathbb{P}(h(X) \neq Y|Y = -1)$ and $R_2(h) = \mathbb{P}(h(X) \neq Y|Y = 1)$ denote a Type I (false positive) and Type II (false negative) error respectively \citep{tong2016survey}. One way to minimize this risk function is to evaluate the conditional probability density functions $f(x|y = +1)$ and $f(x|y = -1)$ from the training set. Assuming that the training sample is representative of the process, then -- in a binary classification system -- it can be shown that the minimized combination of Type I and II errors occurs where these two densities intersect (see \cite{schapire2013boosting} pp. 28 for a discussion).\footnote{In practice the intersection of the conditional densities is cumbersome computationally and the decision threshold is made via Gini Impurity or some other information based metric. We find this particular explanation useful to provide intuition and allow for consistency with common graphics shown in most texts covering hypothesis testing.}
	
	\begin{figure}[H]
		\centering
		\caption{Hypothesis Tests as Classification Problems}
		\subfloat{
			\label{fig: adf_stat_ex}
			\includegraphics[width = 1\textwidth, keepaspectratio]{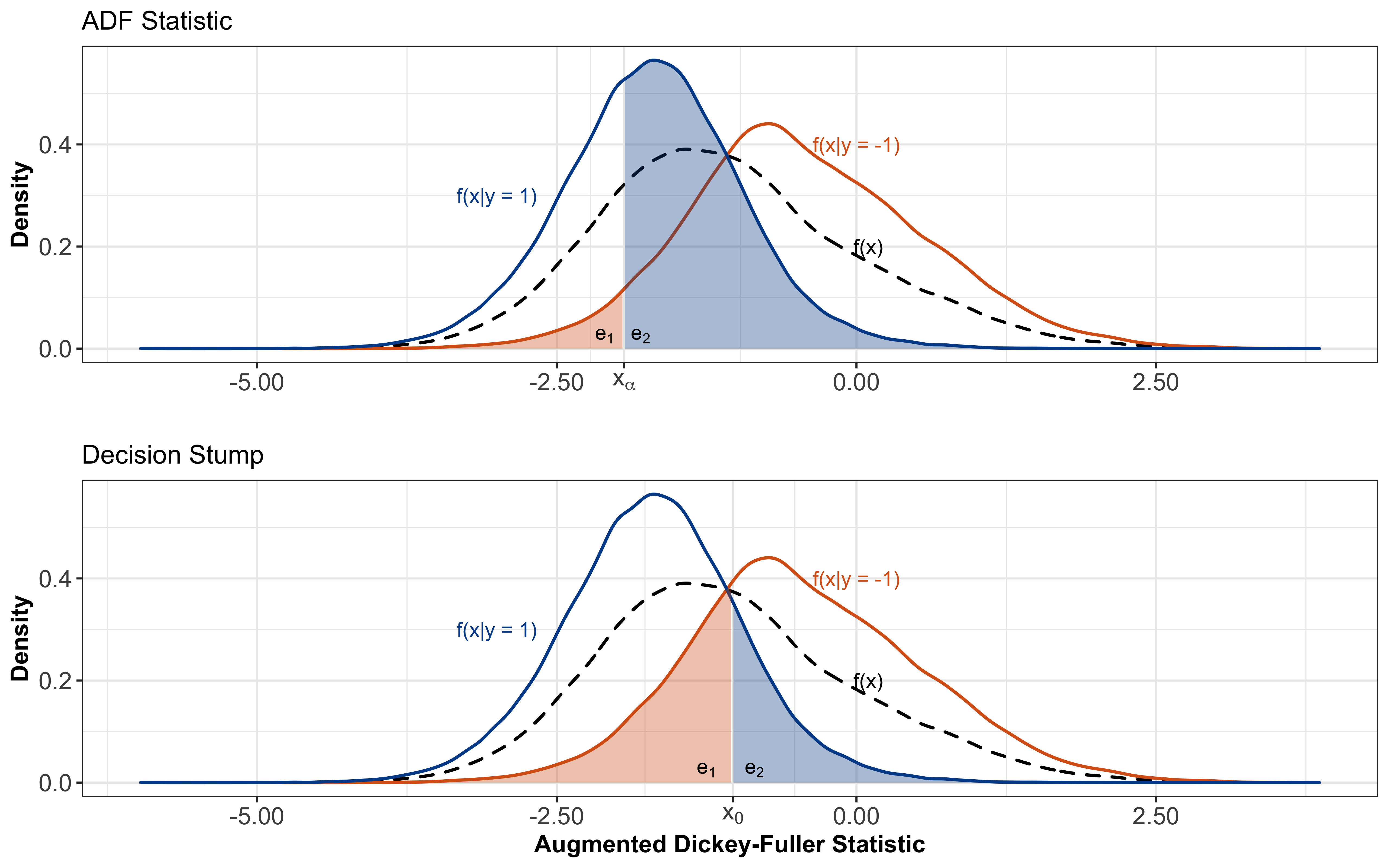} 
		}
	\end{figure}
	
	Let us examine the Augmented Dickey-Fuller test statistic as an illustrative example in Figure \ref{fig: adf_stat_ex}.  In the top panel, we have plotted the ADF statistics, marking the decision threshold with a typical critical value of $\alpha = 0.05$ ($-1.95$). The density of all collected ADF statistics is represented by the dashed line, $f(x)$. Since we have simulated this data we know what the true value of $\phi$ is, and thus we know the true disposition of the series and can outline the resulting conditional distributions $f(x|y = +1)$ when the null is false (\textit{i.e. } the series is stationary), and $f(x|y = -1)$ when the null is true (\textit{i.e. }  the series contains a unit root). Here, we expect a long run Type I error rate of $5\%$ as indicated by our choice of $\alpha$ (shaded area labeled $e_1$), and an indeterminate number of Type II errors (shaded area labeled $e_2$).  
	
	On the other hand, the decision stump has chosen the intersection point between the two conditional densities such that $\zeta \in (-\infty,x_0 \approx -1.03 ]$ and $\zeta^c \in (x_0 \approx -1.03, \infty)$ which leads to a larger Type I error rate but a smaller Type II error rate. As mentioned earlier, this is the point in which the combination of Type I and Type II errors, conditional upon a $0-1$ loss function, is minimized. This means that the overall accuracy of the decision stump is strictly greater than that of the hypothesis test at the chosen $\alpha = 0.05$ for the hypothesis test. Note that both methods rely upon the support of $\mathcal{X}$ which in this case is $\mathbb{R}$. Since $x_\alpha$ and $x_0$ are both chosen over the same support, then it must be the case that for some $\alpha = \alpha'$ the cutoffs chosen are equivalent, $x_\alpha = x_0$. In our example above the decision stump corresponds to a hypothesis test in which $\alpha \approx 0.273$, meaning the hypothesis test $g(x)$, and the decision stump, $h(x)$, are equivalent conditional upon $\alpha = \alpha'$ \citep{neyman1933testing}.
	
	Before we discuss how the Augmented Dickey-Fuller test -- and indeed all hypothesis tests -- can benefit from modern machine learning techniques, we would like to take a moment and show not only an equivalence between the test statistic and decision stump at $\alpha = \alpha'$,  but also that, for any choice of $\alpha$, a boosted statistic can recover the full size-power trade off of a hypothesis test.  
	
	Suppose one would like to apply a supervised learning algorithm to a hypothesis testing problem, but  restrict the long-run Type I error rate to $5\%$. This task can be accomplished through boosting algorithms  \citep{schapire2013boosting}. Boosting extends our single decision stump by combining two or more stumps to improve classification of the phenomenon of interest -- giving the technique the latitude to map the hypothesis space. Let us consider the case of Adaptive Boosting (AdaBoost), a supervised learning algorithm where the learning takes place through an iterative process. In a nutshell, the Adaboost algorithm trains a weak learner (almost always a decision stump), re-weighting the sample based on the mistakes of that learner, then training another stump on a  re-weighted sample.  \footnote{Note that because the weak learner used in the AdaBoost algorithm is the decision stump does not prohibit the boosting of multiple features. At each iteration a different feature may provide a better classifying decision stump based on the re-weighted sample.} Algorithm \ref{algo: boost} provides the basic logic behind AdaBoost.  This has been shown to be surprisingly accurate and relatively robust to overfitting, see \cite{friedman2001elements},\cite{schapire2013boosting}, and \cite{kuhn2013applied} for a more robust discussion regarding the diversity in algorithms, strengths, weaknesses, and implementation.\footnote{For our purposes, in R version 4.0.2 -- ``Taking Off Again" -- we rely on the packages ``Rweka" and ``ada" to implement our decision stumps and boosting algorithms respectively.} 
	

		\begin{algorithm}[H]
			\caption{Adaboost, (\cite{schapire2013boosting}, pp. 5)}

			\begin{algorithmic}
				\State Given $\mathcal{D} = (x_1,y_1),\ldots, (x_m,y_m)$ where $x_i \in \mathcal{X}, y_i \in \{-1,1\}$.
				\State 1. Start with weights $D_1(i) = 1/m$ for $i = 1,\ldots,m$.
				\State 2. Repeat for $t = 1,\ldots, T$:
				\BState \indent (a) Train weak learner (\textit{e.g.} decision stump) using distribution $D_t$.
				\BState \indent (b) Select $h_t$ to minimalize the weighted error: $\epsilon_t = \boldsymbol{Pr}_{i\in D_t}[h_t(x_i) \neq y_i]$.
				\BState \indent (c) Choose $\alpha_t = \frac{1}{2}\text{ln}\big(\frac{1-\epsilon_t}{\epsilon_t}\big)$.
				\BState \indent (d) Update, for $i = 1,\ldots,m$:
				\begin{align}
				D_{t+1}(i) = \frac{D_t(i)\text{exp}(-\alpha_ty_ih_t(x_i))}{Z_t}, \nonumber
				\end{align}
				where $Z_t$ is a normalization factor.
				\State 3. Output $H(x) = \text{sign}\big(\sum_{t = 1}^{T}\alpha_th_t(x)\big).$
			\end{algorithmic}	
		\end{algorithm}\label{algo: boost}
		
\begin{figure}[ht] 
		\centering 
		\caption{Receiver Operating Characteristic Curves: Convergence from Boosting}
		\includegraphics[width=1\linewidth,keepaspectratio]{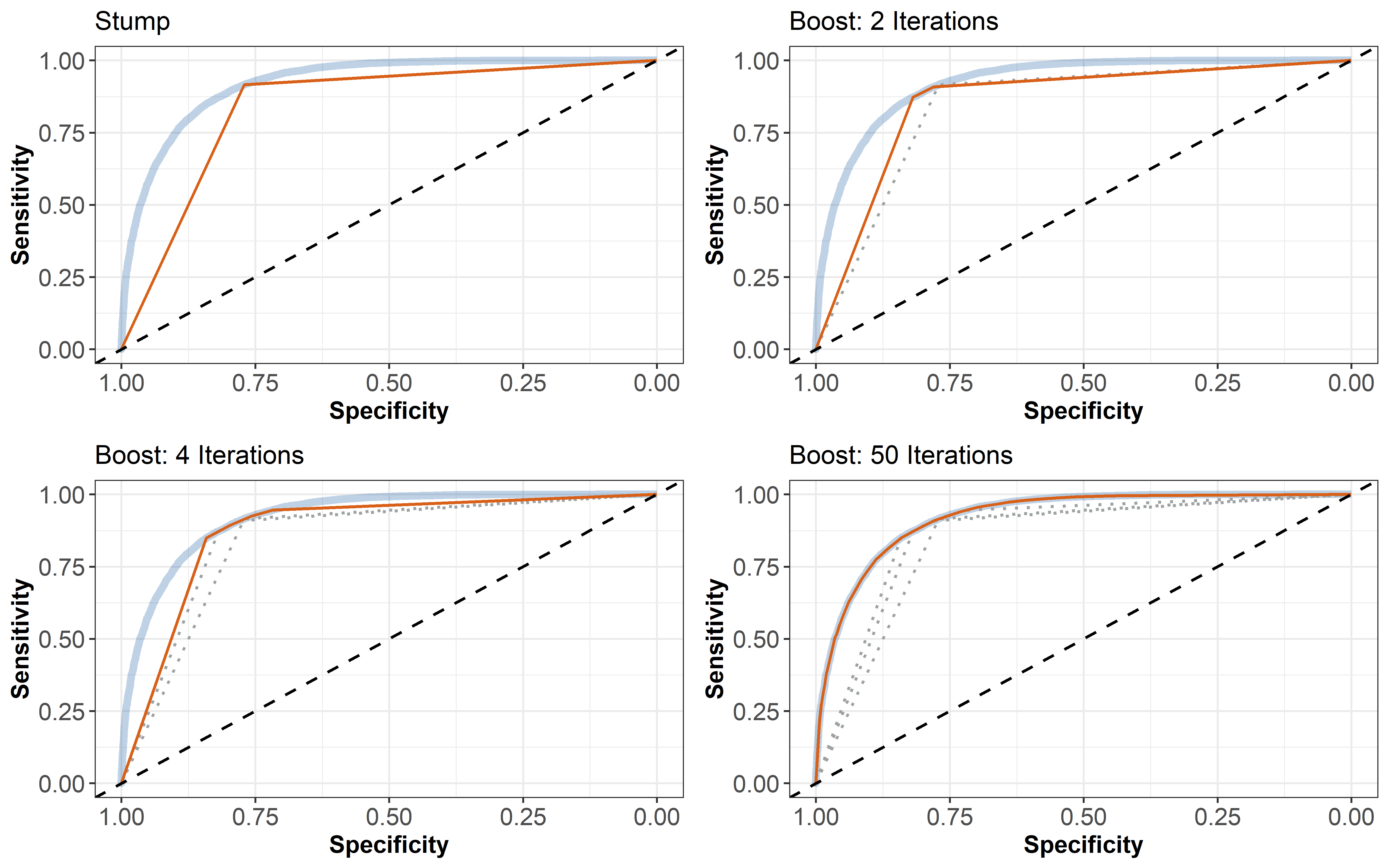} \label{fig: roc_converge}
		\caption*{\footnotesize{\textbf{Note: }Here we are plotting the Receiver Operating Characteristic (ROC) curve under the null (thick light-blue line) versus the boosted alternatives. To calculate the ROC curve under the null we calculated the corresponding probability for the ADF statistic under the cumulative probability distribution of the [simulated] null. This allows for a direct comparison to the predictions of the boosting algorithm.}}
	\end{figure}

	In Figure \ref{fig: roc_converge} we present how boosting converges towards the performance of the ADF statistic with increased number of iterations. The first plot depicts the performance of a single decision stump, which neatly intersects the ADF. When AdaBoost is trained with two iterations, the ROC curve intersects the ADF at two points, that is we have recovered two values of $\alpha'$ for which $\alpha = \alpha'$. In each successive figure, we increase the number of iterations the boosting algorithm is allowed to evaluate. By 50 iterations, the ROC curve for the boosted test statistic has converged to that of the null distribution. It should be clear that any weighted average of cutoffs produced by the boosting algorithm will produce some level of $\alpha' = \alpha$ and thus a boosted test statistic is equivalent to a standard hypothesis test for any choice of long-run Type I errors that may be made. Thus, if we would like to restrict our Type I errors to $\alpha = 0.05$ we can then simply chose a probability threshold such that Specificity over the validation set is $1 - \alpha = 0.95$.
	
	It is important to note that any model or test -- no matter how simple or sophisticated -- is only as good as the data on which it is trained and validated. For an algorithm to retain its predictive qualities in the ``wild”, the training data must reflect the conditions that will be encountered when applied in real world use. In many cases, the qualities of a phenomenon are impossibly expansive to capture in a neat empirical definition, thus the data and  --  in turn the test -- run the risk of biasing from the true class distribution. However, if the definition of a phenomenon is well-specified, one can simulate from the definition (\textit{i.e.} the DGP) so an algorithm can have ample opportunity to thoroughly inspect and distinguish between the null and the alternative.   Unit roots fit into this latter case, we can be certain that our training set accurately reflects the hypothesis test under consideration. Increasing the size of this training set will only improve the accuracy of our threshold choice.
	
	\subsection{A Simple Procedure for Test Construction}
	
	Given the equivalence between hypothesis testing and binary classification problems, we can take full advantage of machine learning algorithms as the mapping function to enhance the power of a test. This can be done for any case in which there is a single test statistic with a direct equivalence for any desired level of Type I error rate, and can include additional information as we will outline shortly. For the Augmented Dickey-Fuller test, and any hypothesis test more broadly we recommend the following:
	
	\begin{enumerate}[noitemsep]
		\item Simulate a balanced training, validation, and test set containing representative cases of the null and alternative hypotheses; 
		\item Derive right-hand side parameters from one or multiple test statistics as well as attributes of the time series of interest; 
		\item Train a set of supervised classifiers, then select the model that fairs the best in cross-validation; then
		\item Given a cost ratio that indicates the importance of  Type I versus Type II errors, calculate the optimal probability threshold from the validation set for classifying individual instances.
	\end{enumerate}
	
	Note that the first step of this procedure requires clarity of purpose, much like in the Neyman-Pearson framework: \textit{a null and its alternative must be clearly defined and replicated through simulation}.  Without a well-defined hypothesis, inference will invariably be confounded by poor and imprecise information.  The second step acknowledges that the role of a mapping function is to represent the hypothesis space as accurately as possible. While a single test statistic embraces parsimony, we advocate for considering all \textit{necessary} information in the form of RHS variables, such as the individual parameters of a current tests (\textit{e.g.} the t-test's parameters are the sample mean and standard deviation), other test statistics which test the same or similar null  hypothesis, and any other correlated variables of interest.  Lastly, a supervised classifier can map the hypothesis space as a function of the RHS variables.  As is common in computer science, we can also conduct a predictive horse race to identify which technique (\textit{e.g.} test statistic, Gradient Boosting, Random Forests, Neural Network, etc.) is the most suitable means of predicting a null or alternative case.

\section{A  ML-Based Unit Root Test}\label{sec: mltest}
	
	In many respects, the hypothesis test's ability to achieve a desired level of performance is rooted in the test's design: \textit{ It is fundamentally a modeling problem}. The current generation of test statistic-based hypothesis tests can be viewed as a simplified approximation of unit roots, combining input parameters in a way that balances practical arithmetic with accuracy. However, applying even a rudimentary ML algorithm to the input parameters of a single test statistic can offer marked gains. In the previous section, we have shown that a simple boosting algorithm applied to the ADF's underlying parameters can achieve the same if not better performance. Thus, the performance of the ADF -- in its current form -- can be considered the lower bound of a ML-based approach.  It  stands to reason that we can achieve improved performance by constructing a composite test that integrates parameters from multiple tests and an arbitrary number of relevant features derived from time series of interest. In this section, we lay out the blueprint of such a ML-based composite test and illustrate the marked performance improves made possible through this novel testing strategy.

	\subsection{Input Features}
	
	The input features are the drivers of predictive accuracy in any modeling problem and are dependent on how much unique information each input feature offers. As we have seen in previous sections, there are marked differences in accuracy test statistics, which can also be viewed as disagreement among these test.  From another perspective, the disagreement is an opportunity  to reconcile seemingly uncorrelated information that can make a test more robust. The question is \textit{How much unique information does each input feature provide a composite ML test?}
	
\begin{figure}[H]
		\centering
		\caption{Variation in Test Statistics.}
		\label{fig: variation}
		\subfloat{
			\includegraphics[width = 1\textwidth,keepaspectratio]{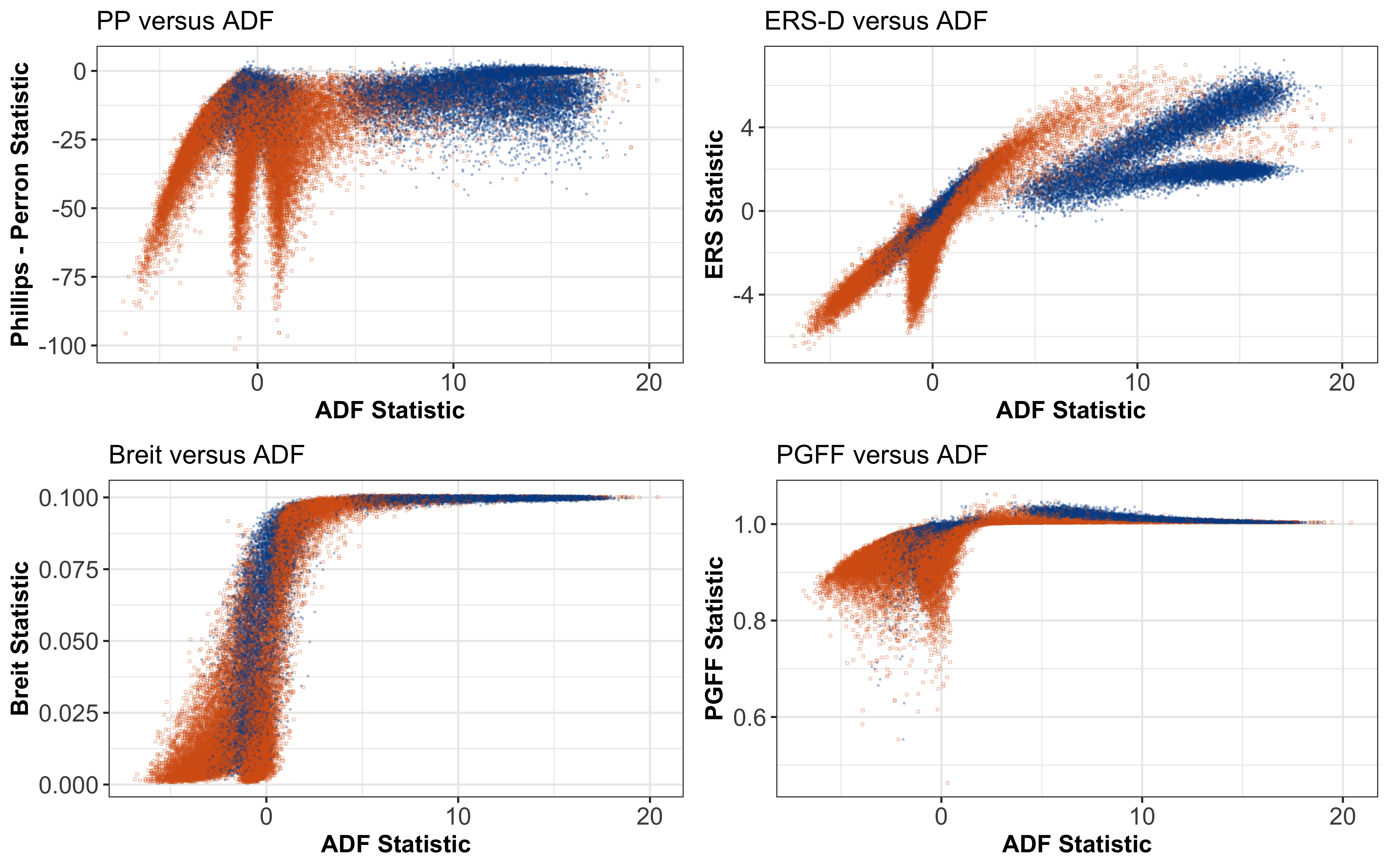}
		}
		
		\caption*{\footnotesize{\textbf{Note:} Blue points are series with a unit root while orange points are stationary series.}}
		
	\end{figure}

	By drawing on our bank of simulated time series, we can examine the quality of information through two-way comparisons between test statistics in Figure \ref{fig: variation}.  Perhaps what is most striking about these bivariate comparisons is that none exhibit a continuous, linear relationship. For example, the ERS-d and the ADF exhibit non-linear and disjoint relationships that could not be captured through linear methods.” The complexity of these patterns is unsurprising as the hypothesis space combines three different unit root processes, requiring one to diagnose each time series for its "true" unit root process in order to identify the appropriate critical value. One could imagine that misidentifying the time series process could lead to misleading inferences, adding to the lack of agreement between tests.  
	
	We can further investigate these relationships between test statistics by calculating their mutual information. While a Pearson's correlation could how quantify the strength of their \textit{linear} relationship, we can evaluate the uniqueness of each test's information through \textit{mutual information}, which is founded in information theory. By calculating how each test encodes information as \textit{bits}, we can estimate mutual information, given as: 
	\begin{align}
		I(X;Y) = \sum_{x \in X} \sum_{y \in Y}  P(x,y) log \frac{P(x,y)}{P(x)P(y)}. \label{eq: mutualinfo}
	\end{align}
	Mutual Information establishes how much information -- linear or non-linear -- is shared between two random variables, which in turn gives a sense of how unique signal is present in among our input features. Further, one can look at the \textit{Information Quality Ratio} defined as,
	\begin{align}
		\mathrm{IQR}(X,Y) = \mathbb{E} \big[ I(X;Y) \big] = \frac{I(X;Y)}{H(X,Y)}, \label{eq: informationqualityratio}
	\end{align}
	where $H(X,Y)$ is the joint entropy of the two random variables. The Information Quality Ratio measures the amount of information in $X$ based on $Y$ against complete uncertainty \citep{wijaya2017information}. In Table \ref{tbl: mutual_information}, we present the Information Quality Ratio for each pair of tests where a value of $1.0$ indicates observing random variable $X$ provides perfect information about random variable $Y$, or more plainly; there is no unique information in $Y$ relative to $X$ . A value of $0.0$ indicates that observing $X$ provides no information about $Y$, each random variable is unique in its information content. In each pairwise comparison, at least 70\% of the information is unique. 
	
		\begin{table}[H]

\caption{Mutual Information of Test Statistics}
\label{tbl: mutual_information}
\centering
\resizebox{\linewidth}{!}{
				\begin{tabular}[t]{lccccccccc}
				\toprule
				  & ADF & KPSS & PP & PGFF & Breit & ERS-d & ERS-p & URZA & URSP\\
				\midrule
				ADF & 1.000 &  &  &  &  &  &  &  & \\
				KPSS & 0.158 & 1.000 &  &  &  &  &  &  & \\
				PP & 0.084 & 0.028 & 1.000 &  &  &  &  &  & \\
				PGFF & 0.142 & 0.142 & 0.064 & 1.000 &  &  &  &  & \\
				Breit & 0.218 & 0.214 & 0.052 & 0.131 & 1.000 &  &  &  & \\
				ERS-d & 0.273 & 0.162 & 0.101 & 0.144 & 0.259 & 1.000 &  &  & \\
				ERS-p & 0.085 & 0.068 & 0.084 & 0.027 & 0.037 & 0.064 & 1.000 &  & \\
				URZA & 0.069 & 0.026 & 0.206 & 0.041 & 0.038 & 0.073 & 0.092 & 1.000 & \\
				URSP & 0.061 & 0.024 & 0.257 & 0.056 & 0.044 & 0.081 & 0.043 & 0.096 & 1.000\\
				\bottomrule
				
				\end{tabular}}

		\caption*{\footnotesize{\textbf{Note:} In this table we have provided the Information Quality Ratio (IQR) for each two-by-two test comparison. The IQR measures the amount of mutual information, $I(X;Y)$, relative to the joint entropy $H(X,Y)$ and is a representation of total correlation.  }}
		
\end{table}

	Given what we have exposed about the interface between near unit roots and unit roots, we can imagine that mapping the hypothesis space may require more information than what is available in current unit root tests. Because ML techniques can evaluate and integrate an arbitrarily large number of input features, we believe that saturating the model with additional input features can more richly represent  the contours of the unit root surface. While traditional statistical and econometric methods run the risk of unstable results when over-parameterized, ML techniques are engineered to mitigate the effects of overfitting.   As outlined in Table \ref{tbl: features}, we include a set of meta-characteristics calculated from each time series as outlined by \cite{wang2006characteristic} and \cite{wang2009rule}. Moreover, we include the series length, frequency, and variance ratio between the first difference and level data.\footnote{The variance ratio is a heuristic whereby evidence of non-stationarity in the level series is present if the ratio between var($\Delta y$)/var($y$) is less than one half.} 
	

		\begin{table}[H]
			\centering
			\caption{Features for Classification}
			\label{tbl: features}
			\begin{tabular}{llll}
				\toprule
				UR Tests & Level and First Difference & STL Decomposed Series & Miscellaneous            \\ \midrule
				ADF      & Skewness                   & TNN Test              & Length                   \\
				PP       & Kurtosis                   & Skewness              & Frequency                \\
				PGFF     & Box Statistic              & Kurtosis              & var($\Delta y$)/var($y$) \\
				KPSS     & Lyapunov Exponent          & Box Statistic         &                          \\
				ERS  (d \& p)      & TNN Test                   &                       &                          \\
				URSP     & Hurst Exponent             &                       &                          \\
				URZA     & Strength of Trend          &                       &                          \\
				Breit    & Strength of Seasonality    &                       &                          \\ \bottomrule
			\end{tabular}
			\caption*{\footnotesize{\textbf{Note:} The statistics calculated on the STL decomposed series are done both on the level and the first difference. Going forward we use $\Delta$ to denote a statistic on the first difference and ``Decomposed" to denote those on the adjusted series. TNN Test refers to the Teraesvirta Neural Network test. See \cite{wang2006characteristic}, \cite{wang2009rule}, and \url{https://robjhyndman.com/hyndsight/tscharacteristics/} for more information.}}
		\end{table}

	When the current generation tests are applied to real world problems, we can now assume that any pair of tests will arrive at seemingly contradictory verdicts about the presence of a unit root tests.   By drawing upon a full spectrum of unit root statistics and features, as many ML-based approaches are effective in handling large feature sets and non-linearities, we believe that even a \textit{standard} set of algorithms can effectively map the contours of the NUR-UR interface and reconcile seemingly divergent information about unit root cases.  
	
	\subsection{ Mapping Functions}
	
	While the center of gravity of the machine learning field has shifted to deep learning, we believe that traditional machine learning can be employed as reliable, no frills mapping functions that serve as useful composite tests.  We thus focus on random forest \citep{breiman2001random} and gradient boosting \citep{friedman2001greedy, chen2016xgboost},  which are both tree-based ensemble learners that are well-suited for structure data problems and can accommodate an arbitrarily large number of variable interactions to map non-linear and disjoint hypothesis spaces. 
	
	A Random Forest (RF) is an ensemble method that uses three main ideas to build better predictions; decision trees, random feature sampling (or random subspace method), and bootstrap sampling \citep{breiman2001random}.  Let $\mathcal{D} = \{(\boldsymbol{x}_1, y_1), \ldots, (\boldsymbol{x}_m,y_m\}$ be a set of training data where $\boldsymbol{x}$ is a $K$-dimensional vector of features specific to each observation and $y \in C$ its corresponding class. It is assumed that $C$ is finite, and in this case we will further assume that $C \in \{-1,+1\}$. It is assumed that each $x_m^k$ is a realization from random variable $\mathcal{X}^k$ having support $\mathbb{R}$ and that these characteristics are measurable features of the data (\textit{e.g.} mean, median, kurtosis, etc.) and a collection of test statistics which are pertinent to some feature of interest (\textit{e.g.} the presence of a unit root or population distribution equality). 
	
	The RF is built from individual decision trees that are grown following a few modifications to standard Classification and Regression Tree (CART) algorithms. In the classification context, a tree splits a sample into smaller, more homogeneous partitions called nodes. Each split is the result of a search -- employing an attribute test --  across features for an optimal threshold, $\theta$, first generating candidate split and selecting the value for $\theta$ that minimizes some loss function. For example, in our case,  the attribute test on which we will rely is Gini Impurity ($G$) which can be written as,
	\begin{align}
		G = 1 - \sum_{c = 1}^{C}p_c^2,
	\end{align}
	where $p_c$ is the proportion of observations which belong to class $c$. In the case of a binary set of classes, such as reject or fail to reject, then $p_c = 0$ if a node contains only observations of the same class, otherwise $p_c > 0$. As one would expect, lower values of $p_c$ indicate greater accuracy. Each child node is further partitioned until some pre-defined stopping criteria are met, \textit{e.g.} minimum node size, or until the models overall Gini Impurity reaches a threshold. The RF algorithm extends this process through bootstrap aggregation and bagging. For each of the $B$ trees that are grown, a bootstrap sample is drawn from the training set, then $k$ input features are drawn and evaluated with an attribute test at each split. When $k$ is small relative to $K$, each tree is "de-correlated" from other trees, producing predictions that are robust to overfitting. In Algorithm \ref{algo: rf} we have outlined the basic \textit{pseudo}-code for the random forest.
	

\begin{algorithm}[H]
	\caption{Random Forest for Classification}
	Input:\\
	Data, $\mathcal{D} = (X,Y)$,  $k$ such that $k<K$,  and a number of bootstrap iterations $B$. \\
	\begin{algorithmic}
		\State 1. For $b = 1$ to $B$:
		\BState \indent (a) Draw a bootstrap sample, $B$, of size $N$ from $\mathcal{D}$.
		\BState \indent (b) Grow a tree, $T_b$, using the bootstrapped sample by recursively repeating the \indent \hspace{1.5em} following until minimum node size is reached:
		\BState \indent \indent (i) Select $k$ at random from the available features.
		\BState \indent \indent (ii) Choose the best variable and split-point from the $k$ options.
		\BState \indent \indent (iii) Split the current node into two child nodes.
		\State 2. Trained model outputs an ensemble of $B$ trees, $\{T_b\}^B$ where the class for each observation, $c_i$, is determined via majority vote by averaging the predictions from each $T_b$. The probability that observation $i$ is in class $C = c$ is $\hat{p}_i = B^{-1}\sum_{b = 1}^{B}R_{mb}$ with determining threshold $\tau$. 
		\vspace{1em}
	\end{algorithmic}	
\end{algorithm}	\label{algo: rf}

	While random forests are a modern classic, a more recent development known as Gradient Boosting has been shown to be a superior approach in Kaggle (and other forecasting) competitions, specifically in the domain of structured data problems \citep{chen2016xgboost}.  Gradient boosting iteratively grows trees to \textit{boost} overall model accuracy. Each iteration is seen as an opportunity to target and correct for the residuals of the previous iteration. All trees are grown  to the same pre-specified terminal depth, but can also be de-correlated from other trees by accommodate bootstrap sampling and random feature sampling to robustify its contributions to the model -- similar to random forests. In general, the algorithm continues to grow additional trees until either a pre-determined number have been formed or until no additional improvement is realized. However, the number of trees is a parameter that is balanced with the learning parameter $\eta$ -- a shrinkage parameter that is designed to minimize overfitting by reducing the contribution of each additional tree. The final prediction is a weighted combination of these iteratively built trees, weighted by the learning parameter.	One special quality of Gradient Boosting algorithms is the architectural differences in the various open source implementations. For this experiment, we rely on Extreme Gradient Boosting (or XGBoost) -- an open source framework for scalable gradient boosting. In Algorithm \ref{algo: gbm} we have outlined the underlying logic of gradient boosted machine \citep{friedman2001greedy}.\footnote{The primary difference between the algorithm presented here and XGBoost specifically is adjustments made for speed and scalability. We encourage readers to see both \cite{friedman2001greedy} and \cite{chen2016xgboost} for more information.}
	

\begin{algorithm}[H]
	\caption{Gradient Boosting for Classification}
	Input:\\
	Data, $\mathcal{D} = (X,Y)$, and a differentiable loss function, $L(y-i,F(x))$. \\
	Initialize model with a $F_0(x) = \underset{\gamma}{\mathrm{argmin}}  \sum_{m}^{M}L(y_m, \gamma)$ \\
	\begin{algorithmic}
		\State 1. For $b = 1$ to $B$:
		\BState \indent (a) Compute $r_{mb} = - \bigg[\frac{\delta L(y_i, F(x_i))}{\delta F(x_i)}\bigg]_{f(x) = F_{b-1}{(x)}} \ \ \text{for} \  m = 1,\ldots,M$
		\BState \indent (b) Fit a tree to the $r_{mb}$ values and create terminal regions $R_{jb}$ for $j = 1,\ldots,J_b$.
		\BState \indent (c) For $j = 1,\ldots,J_b$ compute $\gamma_{jb} = \underset{\gamma}{\mathrm{argmin}} \sum_{x_m\in R_mj} L(y_i, F_{b-1}(x_m)+\gamma)$
		\BState \indent (d) Update $F_b(x) = F_{b-1}(x) + \nu \sum_{j = 1}^{J_b}\gamma_b I(x\in R_{jb})$.
		\State 2. Output $F_B(x)$.
		\vspace{1em}
	\end{algorithmic}	
\end{algorithm}	\label{algo: gbm}

	To identify and calibrate the best mapping function, we perform a \textit{horse race} -- a common practice in computer science, data science, and economic forecasting studies to identify the best model and parameters.  To find the optimal set of hyperparameters for each algorithm, we conduct a grid search for the parameters listed in Table \ref{tbl: hyperparams} following a five-fold cross validation design and using the $M = 350,000$ series training set.  The parameter space spans a total of nine parameter sets for RF and 40 sets for gradient boosting, requiring a total of 245 model runs.  The optimal hyperparameters have been bolded in Table \ref{tbl: hyperparams} . Upon identifying the optimal parameters, we train a "final" version of each algorithm on the full training set, which can then be applied to any new time series to obtain a predicted probability of containing a unit root. 
	

\begin{table}[H]
	\centering
	\caption{Grid Search Parameter Space }
	\label{tbl: hyperparams}
	\begin{tabular}{lcc}
		\toprule
		Hyperparameters       & Random Forest               & Gradient Boosting                          \\ \midrule
		Number of Variables   & $\{3, \boldsymbol{9}, 27\}$ &                                            \\
		Minimum Node Size     & $\{2,\boldsymbol{4},16\}$   &                                            \\
		Eta (Shrinkage Parameter)       &                             & $\{0.01,0.03,\boldsymbol{0.1},0.3,0.5\}$   \\
		Column Sample by Tree  (\% of Columns) &                             & $\{\boldsymbol{0.8}. 1\}$                  \\
		Subsampling of Training Instances           &                             & $\{\boldsymbol{0.8},1\}$                   \\
		Max Tree Depth  (Number of Levels)      &                             & $\{4, \boldsymbol{6}\}$                   \\ \bottomrule
	\end{tabular}
	\caption*{\footnotesize{\textbf{Note}: As identified through five-fold cross validation, the winning value for each hyperparameter is bolded.  }}
\end{table}

	\subsection{Performance of Mapping Functions }
	
	One feature of the ML-based tests is that it should be easily applied to any scenario without modification while test statistics require pre-analysis to identify the type of autoregressive process, which in turn determines the appropriate critical value.  We compute the lower bound performance improvement of the ML-based test by identifying the appropriate critical value for each of the test statistics based on the true simulation parameters. In contrast,  the ML-based tests are applied as if the test set were a real world sample.  We thus calculate  the decision threshold $p_c$ based on an out-of-sample validation set assuming a cost ratio of 1. This threshold is, in turn, applied to the scored out-of-sample test set in order to compute accuracy measures.  As seen in Table \ref{tbl: main_results}, the out-of-sample performance gains from ML-based tests are approximately seventeen percentage points greater than the next highest alternative (ERS-p). Both techniques -- XG and RF -- draw their performance gains from their ability to differentiate near unit roots from unit roots with a sensitivity of 0.924 and 0.925, respectively -- or a thirty-six percentage point improvement over the best test statistic (ERS-p). In contrast, the ML models and traditional test statistics achieve comparable performance in specificity.
	
	\begin{table}[H]

\caption{Main Results}
\label{tbl: main_results}  
\centering
				\scalebox{.85}{
\resizebox{\linewidth}{!}{
\begin{tabular}[t]{lccccccc}
\toprule
  & ACC & SEN & SPE & PPV & NPV & F$^1$ & MCC\\
\midrule
RF & 0.942 & 0.925 & 0.959 & 0.958 & 0.927 & 0.941 & 0.885\\
XG & 0.942 & 0.924 & 0.959 & 0.958 & 0.927 & 0.941 & 0.884\\ \midrule
ADF & 0.763 & 0.546 & 0.980 & 0.964 & 0.684 & 0.697 & 0.583\\
PP & 0.744 & 0.512 & 0.975 & 0.953 & 0.667 & 0.666 & 0.549\\
KPSS & 0.614 & 0.250 & 0.977 & 0.916 & 0.567 & 0.393 & 0.331\\
PGFF & 0.745 & 0.499 & 0.989 & 0.978 & 0.665 & 0.661 & 0.560\\
BREIT & 0.672 & 0.361 & 0.981 & 0.951 & 0.607 & 0.524 & 0.437\\
ERSd & 0.762 & 0.545 & 0.979 & 0.963 & 0.683 & 0.696 & 0.582\\
ERSp & 0.770 & 0.564 & 0.976 & 0.958 & 0.692 & 0.710 & 0.592\\
URZA & 0.635 & 0.309 & 0.959 & 0.883 & 0.582 & 0.458 & 0.354\\
URSP & 0.727 & 0.552 & 0.903 & 0.850 & 0.669 & 0.669 & 0.485\\
\bottomrule
\end{tabular}
}
}
\caption*{\footnotesize{\textbf{Note: } Here we have provided the results from using both the random forest and gradient boosted machine as the mapping function in place of the null distribution. According to both the F$^1$ and MCC measures both are better at classifying a series relative to any single test alternativce. The majority of the improvement comes from increases in sensitivity with both the random forest and gradient boosted machine providing an increase of 36\% over the next best alternative.}}
\end{table}

\begin{figure}[H]
		\centering
		\caption{Power Curves Comparing Test Statistics with ML-Based Test}
		\label{fig: power_roc}
		\subfloat{ 	\includegraphics[width = 1\textwidth,keepaspectratio]{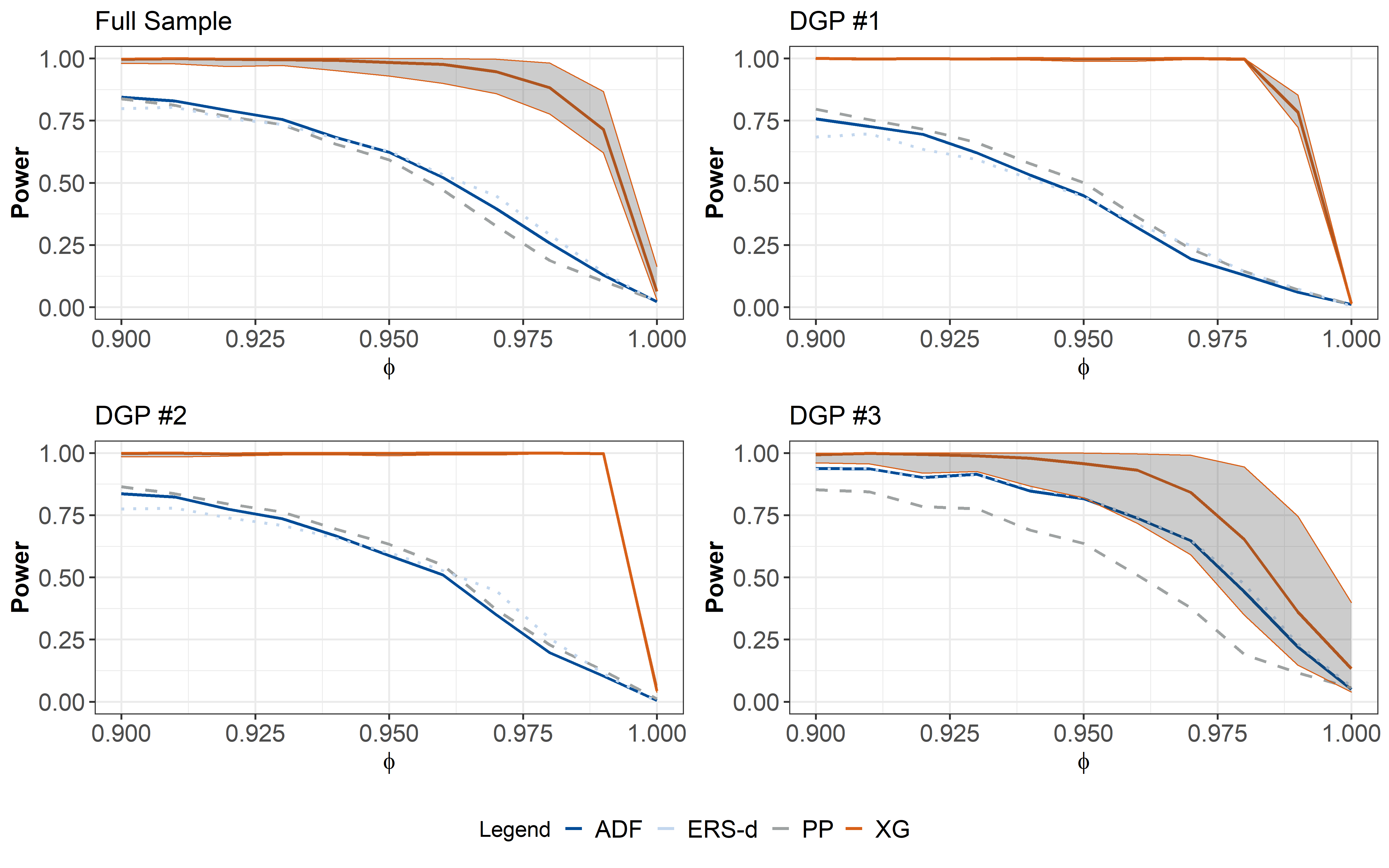} }
		\caption*{\footnotesize{\textbf{Note: } Based on a test sample of $M=75,000$, the power curves are estimated for a cost ratio of 1 at values of $\phi$ with two decimal point precision. The curves correspond to the RF  (orange solid line) as well as the ADF (solid navy blue line), PP (dashed grey line), and ERS-d (dotted light blue line) tests. The upper bound of the RF curve represents performance given a cost ratio of 4 while the lower bound represents a cost ratio of 0.25. }}
	\end{figure}

	Beyond the top-line accuracy, we find evidence that the ML-based approach sustains its relative performance gains across all three DGPs. As shown in Figure \ref{fig: power_roc}, we have constructed four sets of power curves.  The XG-based test has been selected to represent ML techniques given its slightly better performance; However, the difference with the RF-based test is minimal.   The upper bound (the XG evaluated at a threshold set determined by a cost ratio of 4) and the lower bound (cost ratio of 0.25) illustrate how the XG consistently performs above the alternatives.  In the overall sample, the XG-based test comfortably  outperforms the alternatives. However, in the case of DGP \ref{eq: dpg 3} (\textit{i.e.} autoregression without constant or trend), the XG-based test exhibits a greater degree of uncertainty -- the lower bound overlaps with alternative tests, which indicates that the ML option is at least as good as traditional test statistics. Recall however that, for existing tests, we have provided the correct data generating process to the test statistic and thus have provided the upper bound on accuracy. In practice the data generating process is unknown and thus true power would be lower. In contrast, the XG-based test is markedly more effective when applied to more complex autoregressive processes, namely DGP \ref{eq: dpg 1} (\textit{i.e.} autoregression with constant and trend) and DGP \ref{eq: dpg 2}  (\textit{i.e.} autoregression with constant). In fact, the uncertainty is so small that both the upper and lower bounds are nearly indistinguishable from the balanced case. 
	
	To understand what drives these differences in predictive power, we turn to feature importance metrics namely scaled Mean Decrease in Impurity (MDI) -- a measure that estimates the mean reduction to Gini Impurity when the attribute test selects a feature. When scaled, we have a convenient index that identifies the most important feature (\text{i.e.} $100$) and the least important feature (\text{i.e.} $0$). Given the structural differences in each DGP, we isolate the effect of each input feature by re-training the random forest algorithm on the full sample and three subsets, each of which contains one DGP.  The top 10 input features are presented in Table \ref{tbl: variable_importance}. The ADF and the ERS-p both play roles in all samples -- these tests are consistent contributors. However, the Lyapunov Exponent is a core contributor to more complex samples (\textit{e.g.} Full sample, DGPs \#\ref{eq: dpg 1}  and \#\ref{eq: dpg 2}) but not in the simpler sample (\textit{e.g.} DGP \#3). We can infer that the ML algorithm draws on this feature to detect stability within a series, and in turn, identify the discontinuities in the hypothesis space. In a way, inclusion of these statistics help identify the time series process in order to select the appropriate critical value for a test. In this case, the ML streamlines this diagnostic identification process.
	
	\begin{table}[H]

\caption{Feature Importance by DGP}
\label{tbl: variable_importance}
\centering
\resizebox{\linewidth}{!}{
\begin{tabular}[t]{ccccc}
\toprule
 Rank & Full Sample & DGP \#1 & DGP \#2 & DGP \#3\\
\midrule
1 & Lyapunov (100) & Lyapunov (100) & $Z(LX)$: p-val. (100) & ERS-p (100)\\
2 & $Z(LX)$: Std. Err. (100) & ERS-p (100) & ERS-p (87) & $Z(LX)$: Est. (91)\\
3 & ERS-p (96) & Box Statistic (98) & Lyapunov (83) & ADF (85)\\
4 & ADF (95) & Seasonal (85) & ADF (72) & ERS-d (78)\\
5 & $Z(LX)$: p-val. (83) & PP (63) & Trend (63) & $Z(LX)$: p-val. (74)\\
6 & PGFF (66) & ADF (59) & Var. Ratio (57) & PGFF (65)\\
7 & Var. Ratio (61) & Breit (55) & Breit (57) & Var. Ratio (59)\\
8 & PP (59) & Var. Ratio (49) & PGFF (53) & $Z(LX)$: Std. Err. (53)\\
9 & Trend (58) & $Z(LX)$: p-val. (48) & $Z(LX)$: Est. (46) & Trend (53)\\
10 & $Z(LX)$: Est. (56) & URSP (43) & Seasonal (41) & Hurst (33)\\
\bottomrule
\end{tabular}}
\end{table}

	\subsection{Cost-Informed Performance Evaluation}		
	
	Since PPV and Sensitivity are functions of Type I and Type II errors respectively the F-measure can be used to represent asymmetric loss functions. Another way to represent an asymmetric loss function is to change the threshold under which a class is determined. For example, a naive threshold might be $\tau = .5$ indicating that if the simple majority of predictions from each weak learner indicates $y = +1$ then we will assign that class to that observation. It may be the case however that we want more evidence one way or the other with respect to the classification of each observation, such as in the case of unit roots. The rule of thumb with respect to tests for unit roots is that they generally do not have a significant level of power and when in doubt, one should assume non-stationarity and take a first difference. Informally this points to an asymmetric loss function where failing to recognize a unit root when one exists is significantly more costly than the alternative.
	
	\begin{table}[H]

\caption{Results: Cost Ratio Based Threshold Choice}
\label{tbl: cost weights}
\centering
\resizebox{\linewidth}{!}{
\begin{tabular}[t]{lcccccccc}
	\toprule
	           & Threshold & ACC   & SEN   & SPE   & PPV   & NPV   & F$^1$ & MCC   \\ \midrule
	RF$^{4}$ & 0.762     & 0.921 & 0.967 & 0.876 & 0.886 & 0.964 & 0.925 & 0.846 \\
	RF$^{2}$ & 0.625     & 0.938 & 0.950 & 0.927 & 0.928 & 0.949 & 0.939 & 0.877 \\
	RF$^{1}$   & 0.484     & 0.942 & 0.925 & 0.959 & 0.958 & 0.927 & 0.941 & 0.885 \\
	RF$^{.50}$   & 0.356     & 0.937 & 0.898 & 0.976 & 0.974 & 0.905 & 0.934 & 0.877 \\
	RF$^{.25}$   & 0.240     & 0.928 & 0.867 & 0.988 & 0.987 & 0.882 & 0.923 & 0.862 \\ \midrule
	XG$^{4}$ & 0.782     & 0.924 & 0.965 & 0.882 & 0.891 & 0.962 & 0.927 & 0.850 \\
	XG$^{2}$ & 0.671     & 0.936 & 0.951 & 0.921 & 0.923 & 0.950 & 0.937 & 0.873 \\
	XG$^{1}$   & 0.476     & 0.942 & 0.924 & 0.959 & 0.958 & 0.927 & 0.941 & 0.884 \\
	XG$^{.50}$   & 0.322     & 0.937 & 0.896 & 0.977 & 0.975 & 0.904 & 0.934 & 0.876 \\
	XG$^{.25}$   & 0.208     & 0.928 & 0.867 & 0.988 & 0.987 & 0.881 & 0.923 & 0.862 \\ \bottomrule
\end{tabular}}
\caption*{\footnotesize{\textbf{Note: } Using the Receiver operating Characteristic curve we are able to specificy a threshold corresponding to a particular ratio of error costs, $c(e_2)/c(e_1)$. A value of $c(e_2)/c(e_1) = 1$ implies that a Type I error, rejecting the null when the null is true, is of equal weight to a Type II error, failing to reject the null when it is false. If $c(e_2)/c(e_1) < 1$ then a Type I error costs more than a Type II while the reverse is true if $c(e_2)/c(e_1) > 1$.}}
\end{table}

\begin{figure}[H]
		\centering
		\caption{Comparison of Test Size and Cost Ratios}
		\label{fig: alpha_cost}
		\subfloat{ 	\includegraphics[width = 1\textwidth,keepaspectratio]{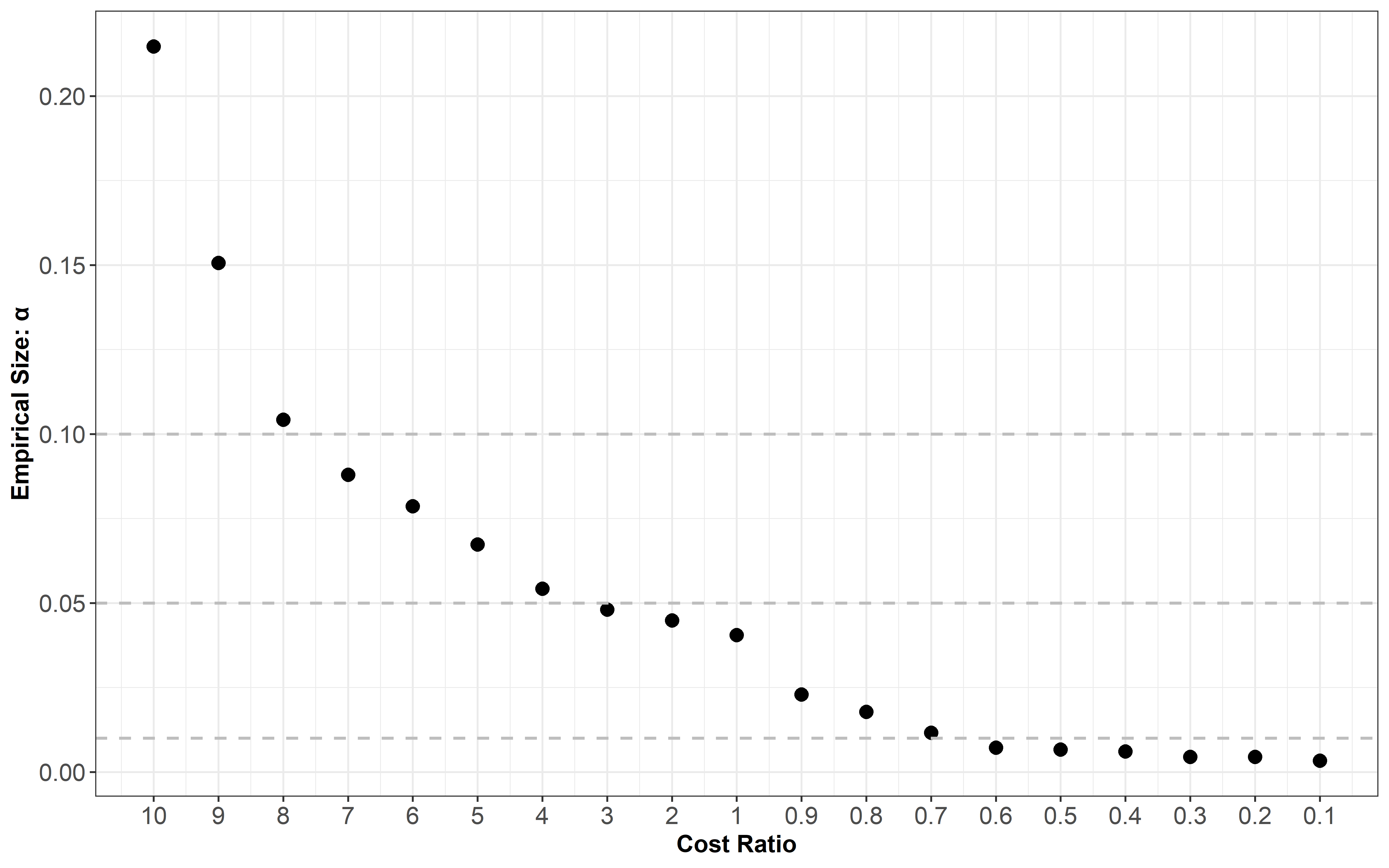} }
		\caption*{\footnotesize{\textbf{Note: }In this figure we are plotting the empirical size of composite unit root test put forth through the gradient boost mapping function with that of a stated cost ratio, $c(e_2)/c(e_1)$. The dashed lines indicate standard choices of $\alpha \in \{0.10, 0.05, 0.01\}$. A choice of $\alpha = 0.05$ implies a cost ratio $c(e_2)/c(e_1) \approx 3$.}}
	\end{figure}

	Using the training set one can directly manipulate the threshold by which a classification is determined in an effort to create an asymmetric loss based prediction. For example, while $c(e^1)$ and $c(e^2)$ may not be known individually, it may be the case that the ratio, $c(e^2)/c(e^1)$, is known. If this is the case, then it is possible to choose a threshold which produces a weighted response of the errors along the ROC curve. In Table \ref{tbl: cost weights} we have provided different thresholds under various assumptions of $c(e^2)/c(e^1)$. Changing the threshold in this manner would shift the power curve presented in Figure \ref{fig: power_roc} up or down based on the cost function ratio. Note that by choosing $\alpha = 0.05$ it implies that $\frac{c(e^2)}{c(e^1)} \approx 3$ (plotted in Figure \ref{fig: alpha_cost}); a Type II error is three times more costly than a Type I error. If a practitioner were to choose $\alpha = 0.01$ then the corresponding cost ratio is $c(e2)/c(e1) \approx 7/10$. Note the advantage of stating a cost ratio rather than a fixed Type I error rate is that the relative importance of the errors is made clear to both the researcher and reader.	While, according to the Neyman-Pearson lemma, the uniformly most powerful test is one which minimizes the Type II errors conditional upon a fixed Type I error rate there is little in the way of quantifying those the Type II errors. Specifying a cost ratio is thus more in line with their (Neyman and Pearson) original thoughts \citep{neyman1933testing}.
	
	By simulating data which satisfies both the null and alternative distribution we are able to use modern ensemble methods via gradient boosting and random forests to form a \textit{pseudo}-composite hypothesis test for the presence of a unit root. This composite test is more accurate, by any reasonable measure of classification accuracy, than any single test statistic. Moreover, since the critical values we use are derived from simulations such as this it stands to reason that a trained aggregator such as the random forest could be provided as part of a standard statistical package providing tests for unit roots.\footnote{We are currently developing such a package which will provide trained models and an appropriate function for calculating the features. This will be available via \url{https://github.com/DataScienceForPublicPolicy/hypML} when complete.}
	
	\section{An Empirical Example}\label{sec: beaugoestotown}
	Following the seminal works of \cite{dickey1979distribution} and \cite{dickey1981likelihood} a conversation began about the presence of unit roots in macroeconomic indicators. \cite{nelson1982trends} examined fourteen such indicators for the presence of a unit root; these included four measures of Gross National Product (real, nominal, per capita, and the deflator), both national employment and unemployment, real and nominal wages, money stock and velocity, bond yields, stock prices, and indices of industrial production and consumer prices. These series were annual in nature and ranged from 62 to 111 years in length. Table \ref{tab:NPsumstat} provides the summary statistics for the data set as put forth by the authors.
	
	\begin{table}[H]
\centering
\caption{Summary Statistics for Nelson and Plosser (1982) Data}
\label{tab:NPsumstat}
\resizebox{\linewidth}{!}{
\begin{tabular}{l|rrrrrr}
Variable              & Start &  End &   T &    Min &     Max &         SD \\ \hline\hline
Real GNP              &  1909 & 1970 &  62 & 116.80 &  724.70 &     180.32 \\
Nominal GNP           &  1909 & 1970 &  62 & 33,400 & 974,126 & 252,334.20 \\
GNP Per Capita        &  1909 & 1970 &  62 &  1,126 &   3,577 &     726.59 \\
Industrial Production &  1860 & 1970 & 111 &   0.90 &  110.70 &      27.65 \\
Employment            &  1890 & 1970 &  81 & 21,102 &  81,815 &  16,755.98 \\
Unemployment Rate     &  1890 & 1970 &  81 &   1.20 &   24.90 &       5.56 \\
GNP Deflator          &  1889 & 1970 &  82 &  22.10 &  135.30 &      31.38 \\
CPI                   &  1860 & 1970 & 111 &  25.00 &  116.30 &      23.41 \\
Nominal Wages         &  1900 & 1970 &  71 &    487 &   8,150 &   2,134.63 \\
Real Wages            &  1900 & 1970 &  71 &  19.48 &   70.81 &      16.46 \\
Money Stock           &  1889 & 1970 &  82 &   3.60 &  401.30 &     102.67 \\
Velocity of Money     &  1869 & 1970 & 102 &   1.16 &    5.61 &       1.14 \\
Bond Yields           &  1871 & 1970 & 100 &   3.14 &   98.70 &      24.05 \\
Stock Prices          &  1900 & 1970 &  71 &   2.43 &    7.60 &       0.96 \\ \hline
\end{tabular}}
\end{table} 

	Of the fourteen indicators included in their analysis the authors came to the conclusion that thirteen of them, the lone exception being the unemployment rate, contained a unit root. As research into unit roots continued over the subsequent decades this data was revisited a number of times including works such as \cite{perron1989great, stock1991confidence, kwiatkowski1992testing, andrews1994approximately,zivot2002further}, and \cite{charles2012trends}. Each of these re-examined the data from \cite{nelson1982trends} using either new tests, assumptions, estimation methods, or a combination of all three, and found as few as three indicators containing a unit root \citep{perron1989great} to a full affirmation of the original work \citep{stock1991confidence,kwiatkowski1992testing,andrews1994approximately}.
	
\begin{figure}[H]
		\centering
		\caption{Comparison of Findings on Data from \cite{nelson1982trends}}
		\subfloat{
		\label{fig: testres}
		\includegraphics[width=.5\linewidth,keepaspectratio]{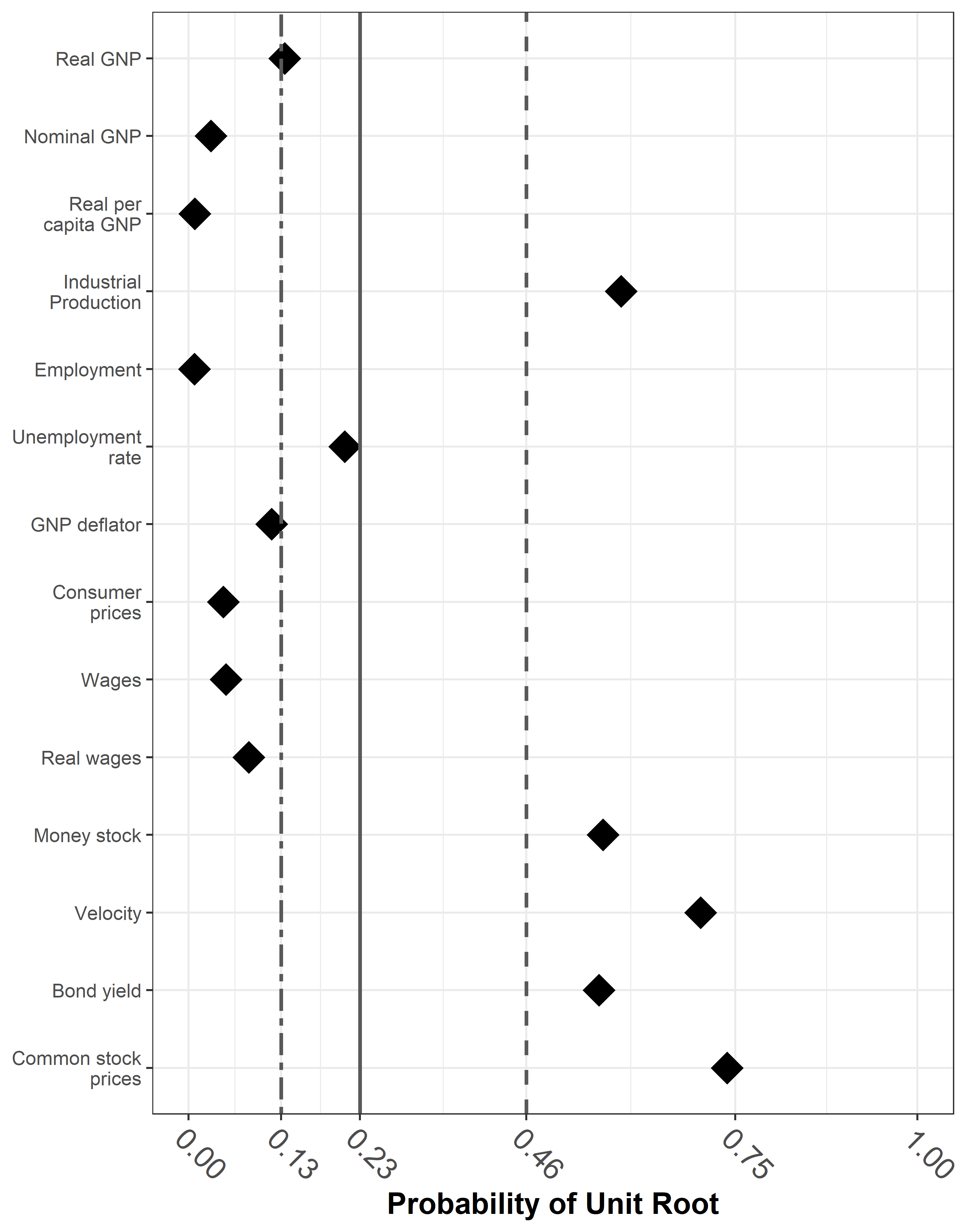}
		} %
		\subfloat{
		\label{fig: comparison}
		\includegraphics[width=.5\linewidth,keepaspectratio]{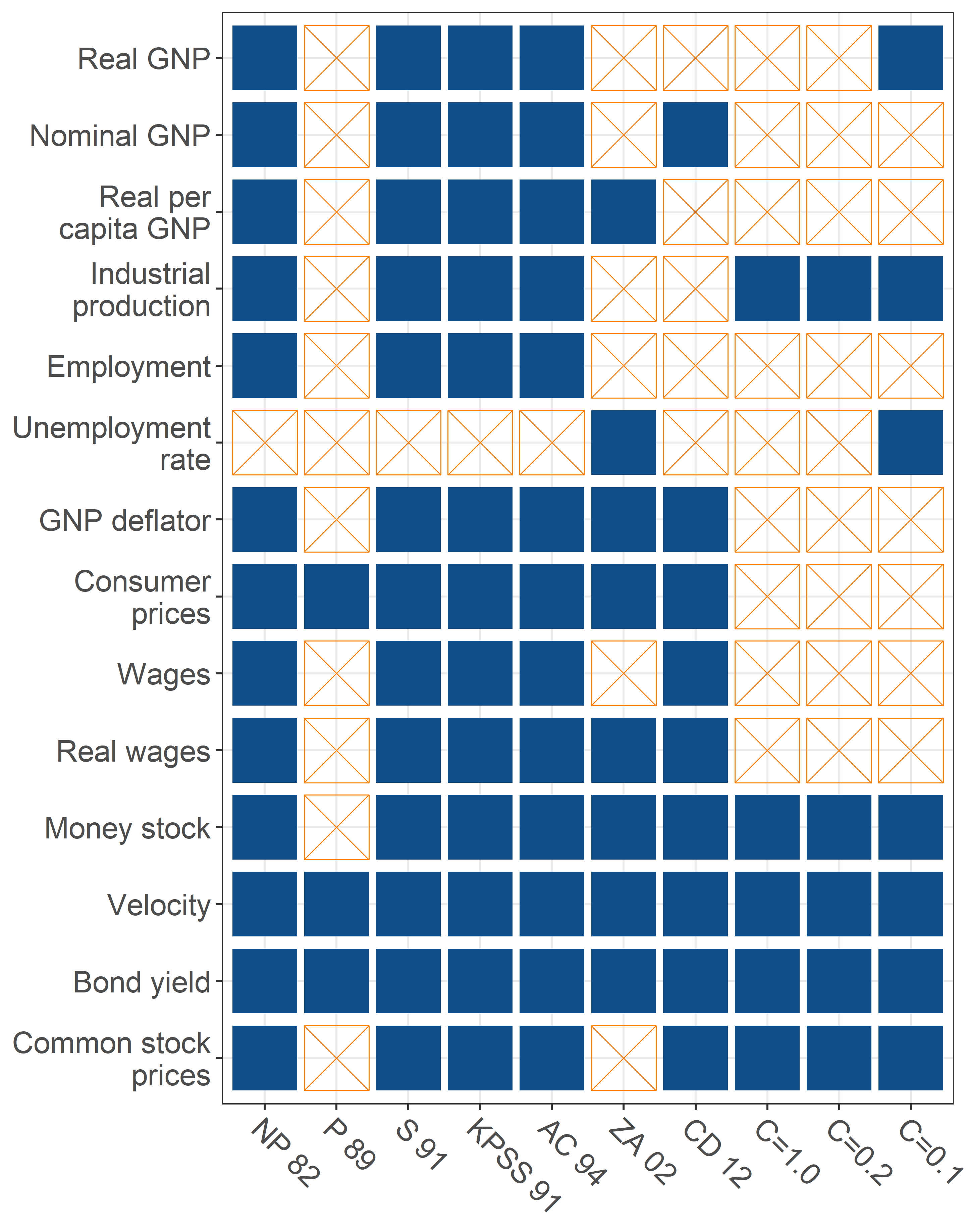}
		}
		\caption*{\footnotesize{\textbf{Note:} Here we have plotted, on the left, the probability each series is a unit root (filled diamonds) in concert with thresholds based on the cost ratios $c(e_2)/c(e_1) = \{1.0,0.20,0.10\}$. For completeness we have also included a threshold such that $\alpha = 0.05$ (light grey dotted line). On the right we have contextualized our results in the literature. If the related paper indicates the series is a unit root it is represented here by a shaded blue square. A decision indicating stationarity is shown here by an orange square with an X inside.}}
	\end{figure}

	Following other authors, we examined the log of these series for the presence of a unit root using our composite testing mechanism. For each series we calculated the full set of features used for training in our mapping function. Further, we assumed three different cost ratios, $c(e_2)/c(e_1) \in \{1.00, 0.20, 0.10\}$, indicating a decision threshold which is progressively more biased towards failing to reject the null. In Figure \ref{fig: testres} we have provided the probability that a unit root is present for each series (filled diamonds) as well as the decision thresholds corresponding to the proposed cost ratios (vertical lines). When presented with a one-to-one cost ratio on both Type I and II errors, the threshold represented by the dashed vertical line, we find that five series -- Industrial Production, Money Stock, Velocity, Bond yield, and Common stock price -- should be considered as indicators containing a unit root. At the more extreme end where Type I errors are ten times more costly than Type II errors, the dot-dash vertical line, an additional two series should be considered a unit root, Real GNP and Unemployment. 
	
	In Figure \ref{fig: comparison} we compare our results to the aforementioned works. Filled squares indicate those series for which the corresponding paper indicated a unit root while the empty square indicates a stationary series. All of the research, including our method at all cost ratios finds that both Velocity and Bond yields should be considered a unit root process. There seems to be broad consensus that Money stock and Common stock prices are also a unit root process. Where we disagree with the majority of previous research is the disposition of Nominal GNP, GNP per capita, GNP deflator, Consumer Prices, Wages, and Real Wages, all of which we find to be trend stationary at even the most favorable of cost ratios. Additionally, we show that, the cost ratio matters when speaking of the Unemployment Rate and Real GNP. If we accept that $c(e_2)/c(e_1) > 1/10$ then both of these series should be considered stationary. However, if we believe that $c(e_2)/c(e_1) \leq 1/10$, rejecting the null of a unit root when true is ten times (or more) costly than failing to reject the null of a unit root when false, then both of these series should be considered a unit root process. 
	
	\section{Conclusion} \label{sec: conclusion} 
	Since its development in the early twentieth century, hypothesis testing has been a foundational tool for making valid statistic inferences from data. In this paper we have updated the hypothesis testing framework by drawing upon modern computational capabilities in the form of machine learning, which not only allows for the creation of pseudo-composite tests that can reconcile the disagreement amongst multiple tests, but enables hypothesis testing to be more in line with the original thinking of \cite{neyman1933ix}. 
	
	By combining common practices in simulation studies and prediction paradigms, our new framework consists of four basic steps. First, simulate a balanced training, validation, and test set which contains representative cases of the null and alternative hypothesis. Second, derive the set of test statistics and other information which can adequately characterize  the hypothesis in question. Third, train a set of supervised classifiers to approximate the complexities of a non-linear hypothesis space and select best performing candidate through cross validation. Finally, given a cost ratio -- $c(e_2)/c(e_1)$ -- classify the optimal probability threshold from the validation set for classifying individual instances. In the event that a suitable cost ratio is not available, or cannot be decided upon, a practitioner can rely on $c(e_2)/c(e_1) = 1$ as a naive choice that minimizes both Type I and II errors.  
	
	To contextualize this new framework we used one of the most visible statistical tests in time series econometrics, that of the unit root. Over the previous four decades a bevy of tests have been constructed to tackle this important question under a variety of circumstances (univariate series, panel data, structural breaks, etc.). Using a simple simulated environment common to the unit root literature we show that the ensemble approach -- using tree-based algorithms that learn from eight common test statistics and other features of the time series bank -- is approximately seventeen percentage points more accurate than the next best, single test alternative. The majority of this increase in accuracy comes from the ability to more accurately reject the null when the null of a unit root when the null is false. The method proposed herein shows a thirty-six percentage point increase in sensitivity (empirical power) relative to the next best, single test alternative without significant size distortion. Traditional power curves, receiver operating characteristic curves, and classification measures (\textit{e.g.} Matthew's Correlation Coefficient) all point to the ensemble method as strictly dominating the evaluated test set.\footnote{We have developed an R package which will be available for practitioners to use our simulated environment to test for the presence of a unit root with no additional computational or programming burden over the standard function syntax for most available test functions.}
	
	Finally, we revisit the fourteen macroeconomic indicators first examined in \cite{nelson1982trends} to place the results of our approach in context with related literature. Of these fourteen indicators the original research and test results indicated thirteen of the fourteen were a unit root process with the lone exception being the national unemployment rate. This specific data set has been revisited over time with as few as three being declared a unit root process \citep{perron1989great}. Our method indicates that, under a balanced cost ratio, only five of these should be considered unit root processes. Under a cost ratio heavily favoring Type I errors, $c(e_2)/c(e_1) = 0.10$, we find an additional two possible unit root processes including unemployment rate.
	
	As we have demonstrated with the unit roots case, the hypothesis space's underlying multivariate distribution are not well-behaved in practice and the abilities of current hypothesis tests are quite variable. Thus, the ability for ML-based hypothesis testing to reconcile conflicting diagnostics and realize large performance gains has far-reaching implications -- the quality of inferences can be vastly improved and the disagreement amongst tests are resolved. Much like replicating a sample of DNA for further study,  ML-based testing can be extended to any phenomenon that is well-defined and can be simulated (\textit{e.g.} equality of distributions, normality). An intriguing possibility is the detection of nuanced sub-conditions (\textit{e.g.} stationary, near unit root, unit root, explosive process), which in turn allows for richer characterizations of random variables and lend greater confidence to the quality of inferences and predictions. 
	
\pagebreak{}

\section{Appendix: Additional Tables}
\begin{table}[H]

\caption{Test Performance for series generated from Equation \ref{eq: dpg 1}}
\centering
\resizebox{\linewidth}{!}{
				\scalebox{.85}{
\begin{tabular}[t]{lccccccc}
\toprule
  & ACC & SEN & SPE & PPV & NPV & F$^1$ & MCC\\
\midrule
RF & 0.979 & 0.958 & 1.000 & 1.000 & 0.960 & 0.979 & 0.959\\
XG & 0.979 & 0.958 & 1.000 & 1.000 & 0.960 & 0.978 & 0.959\\
\midrule
ADF & 0.708 & 0.423 & 0.992 & 0.981 & 0.632 & 0.591 & 0.505\\
PP & 0.722 & 0.453 & 0.991 & 0.980 & 0.644 & 0.620 & 0.526\\
KPSS & 0.571 & 0.159 & 0.983 & 0.903 & 0.539 & 0.271 & 0.250\\
PGFF & 0.692 & 0.385 & 0.998 & 0.994 & 0.619 & 0.556 & 0.485\\
BREIT & 0.652 & 0.309 & 0.995 & 0.986 & 0.590 & 0.471 & 0.419\\
ERSd & 0.703 & 0.410 & 0.997 & 0.992 & 0.628 & 0.580 & 0.502\\
ERSp & 0.706 & 0.421 & 0.990 & 0.977 & 0.631 & 0.589 & 0.500\\
URZA & 0.640 & 0.295 & 0.985 & 0.953 & 0.583 & 0.450 & 0.387\\
URSP & 0.763 & 0.540 & 0.987 & 0.976 & 0.682 & 0.695 & 0.589\\
\bottomrule

\end{tabular}}
}

		\caption*{\footnotesize{\textbf{Note: } The results presented are focused on simulated series generated from Equation \ref{eq: dpg 1} in the test set. The ML-based test accuracy are calculated based on a 1:1 cost ratio while the remaining test statistics are based on the nominal five percent critical value.}}
		
\end{table}
 \label{tbl: enders1}
\begin{table}[H]

\caption{Test Performance for series generated from Equation \ref{eq: dpg 2}}
\centering
\resizebox{\linewidth}{!}{
			\scalebox{.85}{
			\begin{tabular}[t]{lccccccc}
			\toprule
			  & ACC & SEN & SPE & PPV & NPV & F$^1$ & MCC\\
			\midrule
			RF & 0.994 & 0.991 & 0.996 & 0.996 & 0.991 & 0.993 & 0.987\\
			XG & 0.993 & 0.991 & 0.995 & 0.994 & 0.991 & 0.993 & 0.985\\ \midrule
			ADF & 0.760 & 0.522 & 0.996 & 0.992 & 0.678 & 0.684 & 0.589\\
			PP & 0.772 & 0.550 & 0.992 & 0.985 & 0.690 & 0.706 & 0.605\\
			KPSS & 0.647 & 0.291 & 1.000 & 1.000 & 0.588 & 0.451 & 0.413\\
			PGFF & 0.773 & 0.543 & 1.000 & 1.000 & 0.689 & 0.704 & 0.612\\
			BREIT & 0.694 & 0.384 & 1.000 & 1.000 & 0.621 & 0.555 & 0.488\\
			ERSd & 0.763 & 0.524 & 1.000 & 1.000 & 0.680 & 0.688 & 0.597\\
			ERSp & 0.777 & 0.551 & 1.000 & 1.000 & 0.692 & 0.710 & 0.617\\
			URZA & 0.633 & 0.315 & 0.947 & 0.854 & 0.583 & 0.460 & 0.338\\
			URSP & 0.705 & 0.544 & 0.865 & 0.799 & 0.657 & 0.648 & 0.432\\
			\bottomrule
			
			\end{tabular}}
}
		\caption*{\footnotesize{\textbf{Note:} The results presented are focused on simulated series generated from Equation \ref{eq: dpg 2} in the test set. The ML-based test accuracy are calculated based on a 1:1 cost ratio while the remaining test statistics are based on the nominal five percent critical value.}}
		
\end{table}
 \label{tbl: enders2}
\begin{table}[H]

\caption{Test Performance for series generated from Equation \ref{eq: dpg 3}}
\centering
\resizebox{\linewidth}{!}{
					\scalebox{.85}{
				\begin{tabular}[t]{lccccccc}
				\toprule
				  & ACC & SEN & SPE & PPV & NPV & F$^1$ & MCC\\
				\midrule
				RF & 0.854 & 0.825 & 0.882 & 0.875 & 0.835 & 0.850 & 0.709\\
				XG & 0.855 & 0.825 & 0.884 & 0.877 & 0.835 & 0.850 & 0.710\\ \midrule
				ADF & 0.821 & 0.691 & 0.951 & 0.934 & 0.755 & 0.794 & 0.665\\
				PP & 0.737 & 0.532 & 0.943 & 0.903 & 0.668 & 0.669 & 0.520\\
				KPSS & 0.624 & 0.301 & 0.948 & 0.853 & 0.575 & 0.445 & 0.326\\
				PGFF & 0.768 & 0.568 & 0.969 & 0.949 & 0.691 & 0.711 & 0.586\\
				BREIT & 0.670 & 0.391 & 0.949 & 0.884 & 0.609 & 0.542 & 0.409\\
				ERSd & 0.820 & 0.700 & 0.940 & 0.921 & 0.758 & 0.796 & 0.659\\
				ERSp & 0.827 & 0.718 & 0.937 & 0.919 & 0.768 & 0.806 & 0.670\\
				URZA & 0.632 & 0.318 & 0.946 & 0.855 & 0.581 & 0.464 & 0.340\\
				URSP & 0.714 & 0.570 & 0.858 & 0.801 & 0.666 & 0.666 & 0.447\\
				\bottomrule

				\end{tabular}}
			}
		
				\caption*{\footnotesize{\textbf{Note:} The results presented are focused on simulated series generated from Equation \ref{eq: dpg 3} in the test set. The ML-based test accuracy are calculated based on a 1:1 cost ratio while the remaining test statistics are based on the nominal five percent critical value.}}
				
\end{table}
 \label{tbl: enders3}

\pagebreak{}
\bibliographystyle{Chicago}
\spacingset{1} 

\end{document}